\documentclass{emulateapj}

\newcommand{\lsun}{\mbox{L$_\odot$}}
\newcommand{\msun}{\mbox{M$_\odot$}}

\newcommand{\mean}[1]{\mbox{$\langle#1\rangle$}} 
\newcommand{\av}{\mbox{$A_V$}} 

\newcommand{\nh}{\mbox{$N_{H_2}$}} 

\shorttitle{Prestellar cores in Perseus, Serpens, and Ophiuchus}
\shortauthors{Enoch et al.}

\begin{document}

\title{The Mass Distribution and Lifetime of Prestellar Cores in Perseus, Serpens, and Ophiuchus}

\author{Melissa L. Enoch\altaffilmark{1,3}, Neal J. Evans II\altaffilmark{2}, Anneila I. Sargent\altaffilmark{3}, Jason Glenn\altaffilmark{4}, Erik Rosolowsky\altaffilmark{5,6}, and Philip Myers\altaffilmark{6}}

\email{MLE: menoch@astro.berkeley.edu}
\altaffiltext{1}{Department of Astronomy, Univ. of California, Berkeley, CA, 94720}
\altaffiltext{2}{The University of Texas at Austin, Astronomy Department, 1 University Station C1400, Austin, TX, 78712-0259}
\altaffiltext{3}{Division of Physics, Mathematics \& Astronomy, California Institute of Technology, Pasadena, CA 91125}
\altaffiltext{4}{Center for Astrophysics and Space Astronomy, 389-UCB, University of Colorado, Boulder, CO 80309}
\altaffiltext{5}{University of British Columbia, Okanagan, 3333 University Way, Kelowna BC V1V 1V7 Canada}
\altaffiltext{6}{Harvard-Smithsonian Center for Astrophysics, 60 Garden St., Cambridge, MA 02138}

\begin{abstract}

We present an unbiased census of starless cores in Perseus, Serpens,
and Ophiuchus, assembled by comparing large-scale Bolocam 1.1~mm
continuum emission maps with \textit{Spitzer} c2d surveys. We use the
c2d catalogs to separate 108 starless from 92 protostellar  cores in
the 1.1~mm core samples from \citet{enoch06}, \citet{young06}, and
\citet{enoch07}.  A comparison  of these populations reveals the initial
conditions of the starless cores.  Starless cores in Perseus have
similar masses but larger sizes and lower densities on average than
protostellar cores, with sizes that suggest density profiles substantially  
flatter than $\rho \propto r^{-2}$.   By contrast, starless cores in
Serpens are compact and have lower masses than protostellar cores; 
future star formation will likely result in lower 
mass objects than the currently forming protostars.
Comparison to dynamical masses estimated from the
NH$_3$ survey of Perseus cores by \citet{ros07} suggests that most of
the starless cores are likely to be gravitationally bound, and thus
prestellar.  The combined prestellar core mass distribution includes
108 cores and has a slope of $\alpha=-2.3\pm0.4$ for $M>0.8~ \msun$.
This slope is consistent with recent measurements of the stellar initial mass
function, providing further evidence that stellar masses
are directly linked to the core formation process.  We place a lower
limit on the core-to-star efficiency of 25\%.  There are approximately
equal numbers of prestellar and protostellar cores in each cloud,
thus the dense prestellar core lifetime must be similar to the lifetime 
of embedded protostars, or 
$4.5 \times 10^5$ years, with a total uncertainty of a factor of two.  
Such a short lifetime suggests a dynamic, rather than 
quasi-static, core evolution scenario, at least at the relatively high 
mean densities ($n>2\times10^4$~cm$^{-3}$) to which we are sensitive.

\end{abstract}

\section{Introduction}

Dense prestellar cores, from which a new generation of stars will form, 
represent a very early stage of the low mass star formation process, 
before collapse results in the formation of a central protostar. 
The mass and spatial distributions of these prestellar cores retain imprints
of their formation process, and their lifetime is
extremely sensitive to the dominant physics controlling their
formation.  It is well established that most of the star formation in
our Galaxy occurs in clusters and groups within large molecular clouds
\citep[e.g.,][and references therein]{ll03}.     Molecular clouds are
known to be turbulent, with supersonic line-widths
\citep[e.g.,][]{me92},  and to have complex magnetic fields that are
likely important to the cloud physics \citep[e.g.,][]{crutch99}.
Understanding the properties of prestellar cores on molecular cloud scales, and
how they vary with environment, provides insight into the global
physical processes controlling star formation in molecular clouds.   

For example, in the classic paradigm of magnetically dominated star
formation \citep{shu87}, the collapse of cores occurs very slowly via
ambipolar diffusion and starless cores should be long lived, with
lifetimes of order $t_{\mathrm{AD}} \sim 10t_{\mathrm{ff}}$ \citep{nak98}, where
$t_{\mathrm{ff}}$ is the free-fall timescale ($t_{\mathrm{ff}} \sim 10^5$ yr for $n \sim
10^5$ cm$^{-3}$,  where $t_{\mathrm{ff}}$ depends on the mean core density $n$:
$t_{\mathrm{ff}} \propto n^{-0.5}$).   Alternatively, if molecular cloud
evolution is driven primarily by turbulence, over-dense cores should
collapse quickly, on approximately a  dynamical timescale, $1-2
t_{\mathrm{ff}}$ \citep{bp03,mlk04}.  Thus, the lifetime of prestellar cores
should be a strong discriminator of core formation mechanisms.
Published measurements of the prestellar core lifetime vary by  two
orders of magnitude, however, from a few times $10^5$~years to
$10^7$~years (see \citealt{wt07} and references therein).

While the properties of star-forming cores depend strongly on the
physical processes leading to their formation, core initial conditions
in turn  help to determine the evolution of newly formed protostars.
One of the most important diagnostics of initial conditions is the
mass distribution of prestellar cores.  In addition to being a
testable prediction of core  formation models, a comparison of the
core mass distribution (CMD) to the stellar initial mass function
(IMF) may reveal what process is responsible for determining stellar
masses \citep[e.g.,][]{myer00}.  

A number of recent studies have found observational evidence  that the
shape of the IMF is directly tied to the core fragmentation process
\citep{ts98,man98,oni02}.   \citet{alves07} find a turnover in the CMD
of the Pipe Nebula at $M\sim 3 \times$ the turnover  in the Trapezium
IMF \citep{muench02}, and  suggest that the stellar IMF is a direct
product of the CMD, with a uniform core-to-star efficiency of
$30\%\pm10\%$.  The mean particle densities of the
extinction-identified cores in Pipe Nebula study ($5\times10^3 -
2\times10^4$~cm$^{-3}$) are lower than those of typical cores traced
by dust emission  ($2 \times 10^4 - 10^6$~cm$^{-3}$\footnote{The mean
  particle density of Bolocam cores is given by $n = 3M/(4\pi R \mu
  m_H)$, where $M$ is the core mass, $R$ the radius, $\mu=2.33$ is the
  mean molecular weight per particle, and $m_H$ the mass of
  Hydrogen.}; \citealt{enoch07}),  however, and the cores may not be
truly prestellar.  In Orion, \citet{nw07} find a turnover in the CMD
of starless SCUBA $850\micron$ cores at $\sim 1.3 \msun$.  Those
authors relate this turnover to a down-turn in the \citet{kroupa02}
IMF at $\sim 0.1 \msun$, and infer a much lower core-to-star
efficiency of 6\%.

Large samples of prestellar cores are important for further addressing
these problems, as is a more reliable separation of prestellar,
protostellar, and unbound starless cores.  We follow \citet{dif07} in
defining starless cores as low mass dense cores without a compact
internal luminosity source, and prestellar cores, at least
conceptually, as starless cores that are gravitationally bound and
will form stars in the future.  For millimeter cores containing a
compact luminous internal source (i.e., an embedded protostar) we
follow \citet{dif07} in terming these protostellar cores, regardless
of whether the final object will be stellar or sub-stellar in nature.
Unlike protostellar cores, which are internally heated by the embedded
source as well as externally by  the interstellar radiation field
(ISRF),  starless cores are heated only externally by the ISRF, with
decreasing  temperatures toward the core center
\citep[e.g.,][]{evans01}.  

Molecular line or extinction surveys often trace relatively low
density material ($10^3-10^4$~cm$^{-3}$), leaving the possibility that
such cores may never collapse to form stars.  In addition, most
previous studies base the identification of protostellar versus
starless cores on near-infrared data, which is not sensitive to the
most embedded protostars, or on low resolution and poor sensitivity
IRAS maps.   The first issue can be remedied by using millimeter or
submillimeter surveys; (sub)mm emission traces dense  ($n\gtrsim
2\times10^4$ ~cm$^{-3}$; \citealt{wt94,enoch07}) material, and
detection at (sub)mm wavelengths tends to correlate with other
indications of a prestellar nature, such as inward motions
\citep{ge00}.  \textit{Spitzer} provides significant progress on the
second issue, with substantially superior resolution and sensitivity
($0.01 L_{\sun}$ at 260~pc; \citealt{harv07a})  compared to IRAS,
making the identification of prestellar cores much more secure.
Recent examples of identifying starless cores include the studies of
\citet{jorg07} and \citet{hatch07}, which utilize SCUBA
850~\micron\ surveys and Spitzer data to distinguish starless from
protostellar cores in the Perseus molecular cloud.

The combination of (sub)mm studies with molecular line observations of
dense gas tracers, which yield a gas temperature and line-width, is a
powerful method for determining the mechanical balance of starless
cores.   While our preliminary operational definition of a prestellar
core will be a starless core that is detected at submillimeter or
millimeter wavelengths, we will examine this issue more closely in
\S~\ref{boundsec}.  Comparison of our data to molecular line
observations, such as the recent GBT NH$_3$ (2,2) and (1,1) survey of
Bolocam cores in Perseus by \citet{ros07} provides a more robust
method of estimating whether cores are gravitationally bound.

We have recently completed large continuum surveys at $\lambda=1.1$~mm
of the  Perseus, Ophiuchus, and Serpens molecular clouds using Bolocam
at the Caltech  Submillimeter Observatory (CSO).  We mapped 7.5
deg$^2$ (140~pc$^2$ at our adopted cloud distance of $d=250$~pc) in
Perseus, 10.8 deg$^2$ (50~pc$^2$ at $d=125$~pc) in Ophiuchus, and 1.5
deg$^2$ (30~pc$^2$ at $d=260$~pc) in Serpens with a resolution of
$31\arcsec$ \citep[][hereafter Papers I, II, and III,
  respectively]{enoch06,young06,enoch07}.   Millimeter emission traces
the properties of starless cores and protostellar envelopes, including
core sizes, shapes, masses, densities, and spatial distribution.   The
1.1~mm Bolocam surveys are complemented by large \textit{Spitzer}
Space Telescope IRAC and MIPS maps of the same clouds from the ``From
Molecular Cores to Planet-forming Disks'' \textit{Spitzer} Legacy
program (``Cores to Disks'' or c2d; \citealt{evans03}).  Combining
these data sets with the 2MASS survey provides wavelength coverage
from $1.25-160\micron$, and  enables us to reliably differentiate
starless cores from those that have already formed embedded
protostars.  The \textit{Spitzer} c2d IRAC and MIPS surveys of each
cloud are described in detail in \citet{jorg06}, \citet{harv06},
\citet{reb07}, \citet{padg07}, and \citet{harv07a}. 

In Paper III, we looked at how the global molecular cloud environment
influences the properties of star-forming cores, by comparing the
1.1~mm  core populations in Perseus, Serpens, and Ophiuchus.    In a
companion paper to this work (Enoch et al. 2008, in prep), we use the
comparison of 1.1~mm and \textit{Spitzer} data to study the properties
of  embedded protostars.  In particular, we examine the bolometric
temperatures and luminosities of Class 0 and Class I protostars in
Perseus, Serpens, and Ophiuchus, the lifetime of the Class 0 phase,
and accretion rates and history for the early protostellar phases.
Here we use the combination of Bolocam 1.1~mm and  \textit{Spitzer}
c2d surveys to probe the initial conditions  of star formation on
molecular cloud scales, and how the properties of starless cores
differ from cores that have already formed protostars.  

In \S\ref{sepsec} we describe the identification of protostellar
cores, including the combination of 1.1~mm and \textit{Spitzer}
infrared (IR) data (\S~\ref{datasec}), identification of candidate
protostars based on their mid- and far-infrared properties
(\S\ref{pidsect}), and the basis on which we determine association
between 1.1~mm cores and candidate protostars (\S~\ref{assocsec}).
The resulting starless and protostellar 1.1~mm core populations for
each cloud are compared in \S\ref{embsl}, including core sizes and
shapes (\S\ref{sizesec}), masses and densities (\S\ref{densec}),
distribution in mass versus size (\S\ref{mvssec}),  relationship to
cloud column density (\S\ref{probsec}), and spatial clustering
(\S\ref{clustsec}).    We calculate the dynamical mass of starless
cores in Perseus using  NH$_3$ observations to determine if the cores
are truly prestellar (\S~\ref{boundsec}).  We combine the three clouds
to produce the prestellar core mass distribution (CMD), which is
discussed in relation to the stellar initial mass function in
\S\ref{slcmdsec}.  Finally, in \S~\ref{lifesec} we estimate the
lifetime of the dense prestellar core phase and discuss implications
for star formation theory.

\section{Separating Starless and Protostellar Cores}\label{sepsec}

To study the initial conditions of star formation as traced by
prestellar cores,  we first must differentiate cores without an
internal source of luminosity (starless cores) from those with an
embedded self-luminous source (protostellar cores).   Protostellar
cores will have lost some mass due to accretion onto the  embedded
protostar, and may be otherwise altered, so that they  are no longer
representative of core initial conditions.   Starless and protostellar
cores can be differentiated using the \textit{Spitzer} c2d surveys, by
identifying infrared sources that may be associated with a given core.
Such candidate protostars are typically visible as point-like objects
in the near- to mid-infrared data.  In the following sections we
describe the merging of the millimeter and infrared data and the
criteria used to determine which  cores are protostellar.

\subsection{Combining Bolocam and Spitzer c2d Data}\label{datasec}

\textit{Spitzer} IRAC and MIPS  maps from the c2d Legacy program cover
nearly the same area as our  Bolocam 1.1~mm maps of  Perseus, Serpens,
and Ophiuchus.  Both  Bolocam and \textit{Spitzer} maps were designed
to cover down to a visual extinction of $\av \gtrsim 2$~mag in
Perseus, $\av \gtrsim 3$~mag in Ophiuchus, and $\av \gtrsim6$~mag in
Serpens \citep{evans03}.   The actual overlap in area between Bolocam
and IRAC maps is shown in  Figure~1 of Papers I, II and III for
Perseus, Ophiuchus, and Serpens, respectively.  Catalogs listing c2d
\textit{Spitzer}  fluxes of all detected sources in each of the three
clouds, as well as near-infrared fluxes for sources that also appear
in the  2MASS catalogs, are available through the  \textit{Spitzer}
database \citep{evans07}.  Thus, we have wavelength coverage from
$\lambda=1.25$ to $1100~ \micron$, utilizing 2MASS ($\lambda=1.25$,
1.65, 2.17~$\micron$), IRAC ($\lambda=3.6$, 4.5, 5.8, 8.0~$\micron$),
MIPS ($\lambda=24$, 70, 160~$\micron$), and Bolocam ($\lambda=1.1$~mm)
data.  Note that $160~ \micron$ flux measurements are not included in
the c2d delivery catalogs due to substantial uncertainties and
incompleteness,  but are utilized here and in Enoch et al. (2008, in prep.)
when possible. 

Basic data papers describe the processing and analysis of the
\textit{Spitzer} IRAC and MIPS maps of Perseus, Serpens,  and
Ophiuchus, and present general properties of the sources in each cloud
such as color-color and color-magnitude diagrams \citep{jorg06,harv06,
  reb07,harv07a}.     In addition, the young stellar object (YSO)
population in Serpens is discussed in detail by \citet{harv07b}.
Here we are most interested in very red  sources that are  likely to
be embedded in the millimeter cores detected with Bolocam.  For the
following we will use the term ``candidate protostar'' in general to
encompass candidate Class~0 and  Class~I objects \citep{awb93,lw84},
although more evolved  sources may be included in this sample as well.
A more detailed study  of the properties of the candidate protostars
themselves is carried out in  a companion paper (Enoch et al., in
prep).

\begin{figure*}[!ht]
\includegraphics[width=7.3in]{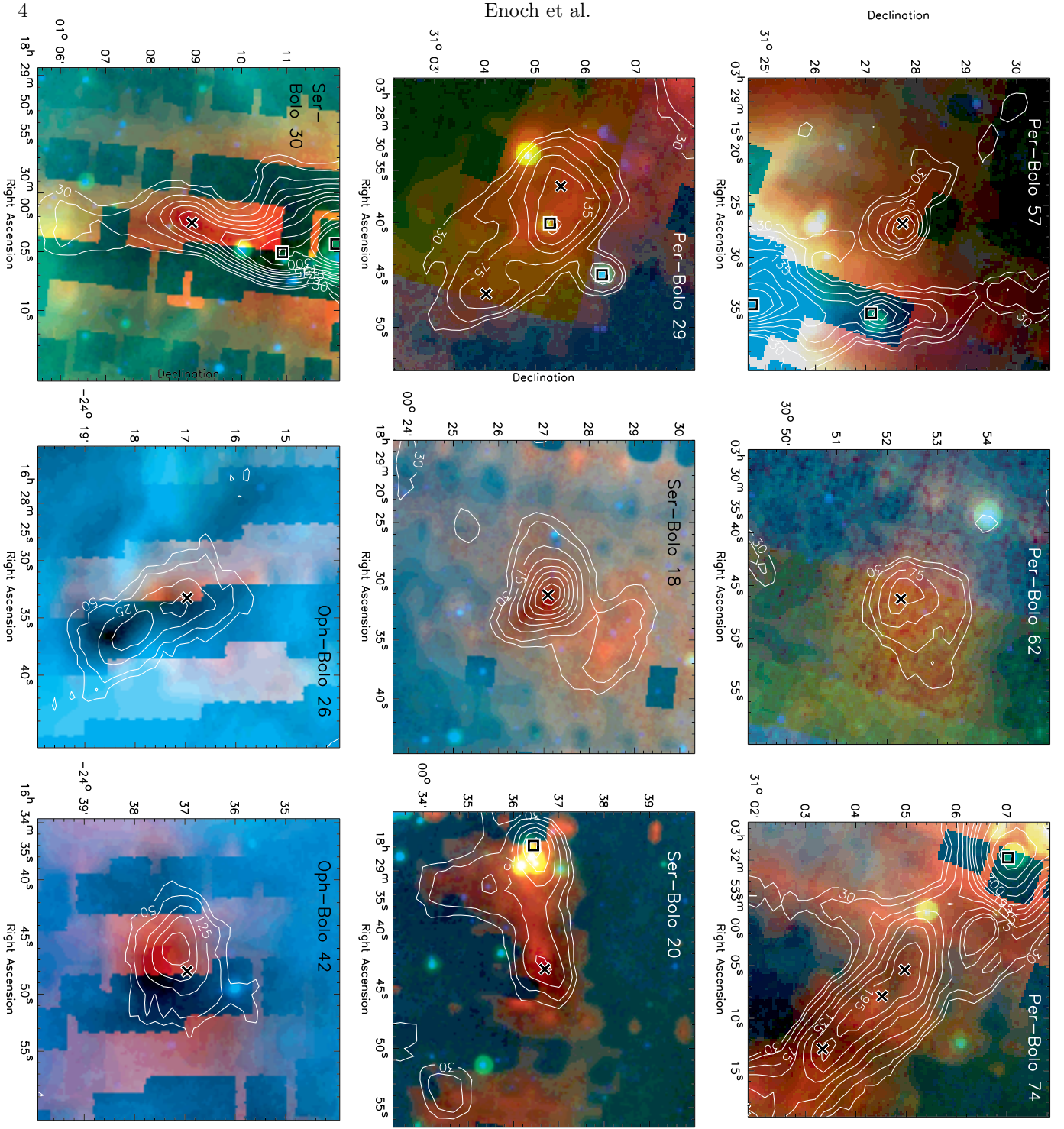}
\caption{Three color Spitzer images (8.0, 24, 160 $\micron$) of
  selected starless and protostellar cores in Perseus, Serpens, and
  Ophiuchus.  Bolocam 1.1~mm contours are overlaid at intervals of
  $2,3,5,...15,20...35 ~ \sigma$, where $1\sigma$ is the mean rms
  noise in each 1.1~mm Bolocam map (15~mJy in Perseus, 10~mJy in
  Serpens, and 25~mJy in Ophiuchus). Positions of starless  cores are
  indicated by an ``x'', protostellar cores by a square,  and the
  Bolocam ID of the centered core is given.   Note the lack of
  infrared point sources near the starless core positions.  Starless
  cores are most notable in  the infrared by their extended
  $160\micron$ emission or dark shadows at shorter  wavelengths, while
  protostellar cores are clearly associated  with \textit{Spitzer}
  point sources.  Saturation, image artifacts, and incomplete sampling
  causes the striped appearance of some of the red $160~ \micron$
  images.  
\label{sl3color}}
\end{figure*}

In Figure~\ref{sl3color} we show the result of combining
\textit{Spitzer} and Bolocam data for a few cores in each cloud.
Three-color (8, 24, 160~$\micron$) \textit{Spitzer} images are
overlaid with 1.1~mm Bolocam contours, and symbols mark the
positions of all identified 1.1~mm starless cores (``x''s) and
protostellar cores (squares).  Bolocam  IDs of the central sources,
from Table~1 of Papers I--III, are given at the top of each image.
Red $160~ \micron$ images are often affected by saturation, pixel
artifacts (bright pixels), and incomplete coverage.  Saturation by
bright sources affects many pixels in a given scan; this and
incomplete sampling accounts for the striped appearance of the red
images, particularly in Ophiuchus.  

Note the lack of infrared point sources near the center of the
starless 1.1~mm cores, whereas protostellar 1.1~mm cores are  clearly
associated with one or more \textit{Spitzer} sources.   Apart from
these common traits, both starless and protostellar cores display a
wide range of properties.  They may be isolated single sources
(e.g. Per-Bolo 62, Oph-Bolo 42), associated with filaments or  groups
(Per-Bolo 74, Ser-Bolo 20), or found near very bright protostars
(Per-Bolo~57).  Although most of the starless cores are extended, some
are quite compact (e.g., Per-Bolo~57), barely resolved by the
31\arcsec\ Bolocam beam.  Some starless cores are distinguished in the
Spitzer bands by their bright $160~ \micron$ emission.  Per-Bolo~62 is
not detectable at 24 or $70~ \micron$, but emerges as a  diffuse
source at $160~ \micron$, indicating a cold, extended core that
closely mirrors 1.1~mm contours.  Other starless cores  stand out as
dark ``shadows'' in the shorter wavelength bands. Ser-Bolo~18 and
Oph-Bolo~26 are two examples of such cores; the dark shadows against
bright 8 and $24~ \micron$ emission are suggestive of dense cores
obscuring  background nebular emission.  Again, contours at 1.1~mm
closely trace the \textit{Spitzer}  short-wavelength shadows.

Although the sky coverage of the IRAC, MIPS and Bolocam maps overlaps
nearly perfectly for our purposes, there is a small portion of the
Serpens 1.1~mm map that is not covered by the $70~ \micron$ map.  In
addition, the $160~ \micron$ maps are often saturated near bright
sources and in regions of bright extended emission, such as near
bright clusters of sources.

\begin{figure*}
\includegraphics[width=7.3in]{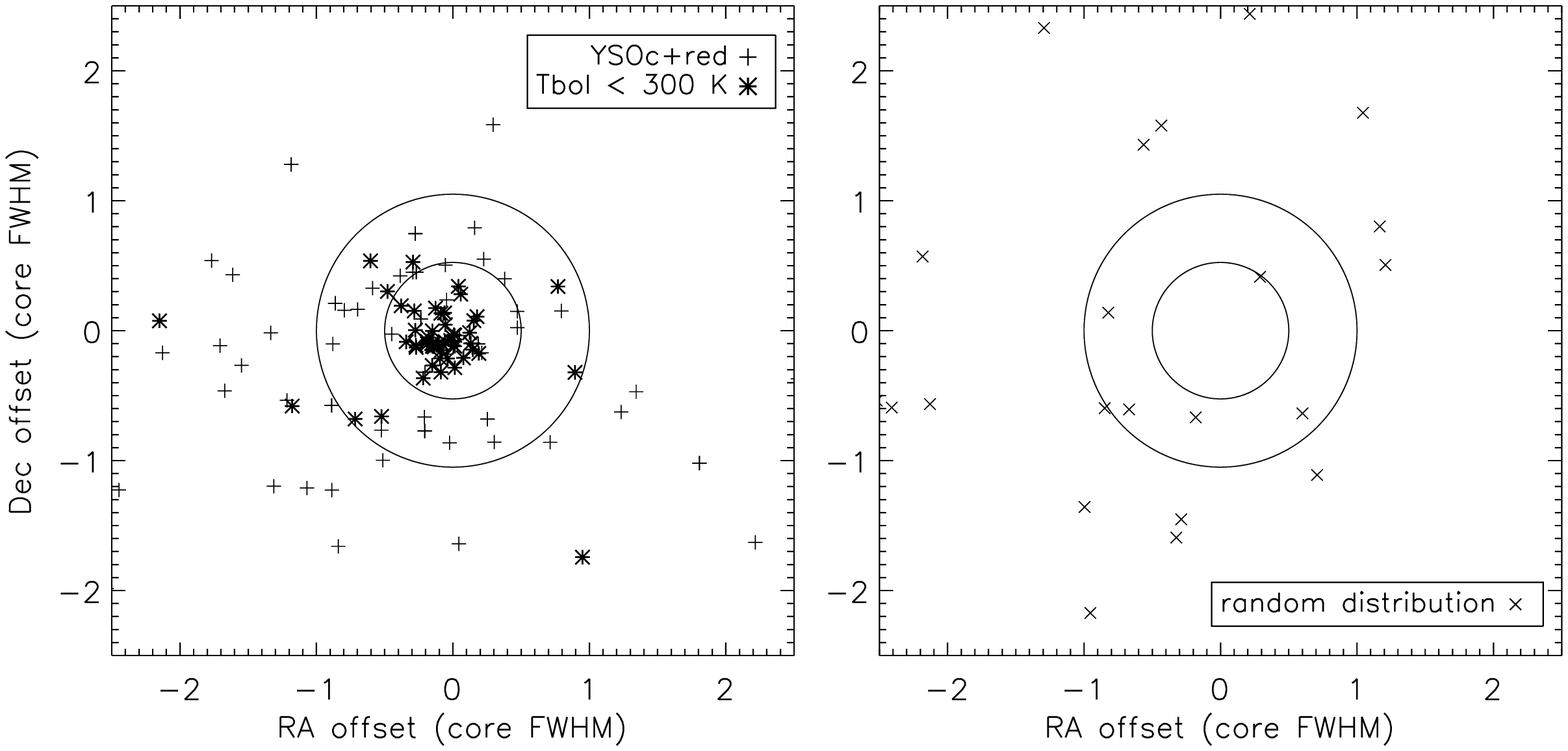}
\caption
{\textit{Left:} Distribution of the positional offset from each
  protostar candidate in Perseus to the nearest 1.1~mm core centroid.
  Distances are in units of the core FWHM size ($\theta_{1mm}$), and
  circles enclose sources within 0.5 and 1.0 $\theta_{1mm}$ of a core
  position.  Pointing errors in the Bolocam maps are typically
  $10\arcsec$ or less.  Protostar candidates include all ``YSOc'',
  ``red'' and $70~ \micron$ sources, while candidates with bolometric
  temperatures $T_{bol}<300$~K (Enoch et al. 2008, in prep) are shown in
  bold.  These cold objects, which are most
  likely to be most embedded, are primarily located within $1.0
  ~\theta_{1mm}$ of a core centroid.  \textit{Right:} Expected
  distribution if the IR sources were distributed randomly over the
  Bolocam map area (i.e. if they were not spatially correlated with
  the cores).   Our definition that a given 1.1~mm core is
  protostellar if there is a candidate protostar within 1.0
  $\theta_{1mm}$ of the core position will result in a few false
  associations in each cloud.  
}\label{jdistfig}
\end{figure*}

\subsection{Identifying Candidate Protostars}\label{pidsect}

The identification of candidate protostars is discussed in more
detail in Enoch et al. (2008, in prep); here we briefly describe the
sample used to identify protostellar cores.  The
first criteria used to select candidate protostars from the c2d
source catalogs is based on the source ``class.''   All sources in the
c2d database are assigned a class parameter  based on colors,
magnitudes, and stellar SED fits, as discussed  in the c2d Delivery
Document \citep{evans07} and in \citet{harv07a}.   Class parameters
include ``star'',  ``star+disk'', ``YSOc''  (young stellar object
candidate), ``red'',  ``rising'', ``Galc'' (galaxy candidate), etc.
Protostar candidates will generally be a subset of YSOc sources, but
some  of the most embedded may also be assigned to the ``red'' class
if  they are not detected in all four IRAC bands.

A preliminary sample is formed from sources classified as ``YSOc'' or
``red.''   We impose a flux limit at $\lambda = 24\micron$ of
$S_{24\micron} \ge 3$~mJy\footnote{Note that this limit is
considerably higher than the $S_{24\micron} \gtrsim 0.7$~mJy limit
imposed by \citet{harv07b}, together with several other criteria, 
to help eliminate galaxies.},  high enough to
eliminate most extragalactic interlopers and sources with SEDs that
are clearly inconsistent with  an embedded nature (e.g., sources with
$S_{24\micron} < S_{8\micron}$), but low enough to include most deeply
embedded known protostars.  In addition, we include any $70~ \micron$
point  sources that are not classified as galaxy candidates
(``Galc'').  In each cloud, a few known deeply embedded sources that
have strong $70~ \micron$ emission but very weak $24~ \micron$
emission (e.g., HH211 in Perseus) are recovered by this last criteria,
as are a few very bright sources that are saturated at  $24~ \micron$
(these are often classified as ``rising'').  The \textit{Spitzer} c2d
surveys are complete to young objects with
luminosities as low as $0.05 \lsun$ (Dunham  et al. 2008, in prep.;
\citealt{harv07a}), and we are unlikely to be missing any protostellar
sources down to this level (Enoch et al. 2008, in prep.;  Dunham et al.
2008, in prep.).

\subsection{Determining association between cores and candidate protostars}\label{assocsec}

We next determine which of the 1.1~mm cores are associated with a
candidate protostar.  The positional offset from each  candidate
protostar to the nearest 1.1~mm core  centroid position, in units of
the core  full-width at half-maximum (FWHM) size, is plotted in
Figure~\ref{jdistfig} (left).  Note that this analysis is similar to
Figure~2 of \citet{jorg07}.  Large circles enclose candidate
protostars  that are located within $0.5\times \theta_{1mm}$ and
$1.0\times \theta_{1mm}$ of a 1.1~mm core position, where
$\theta_{1mm}$ is the angular FWHM size of a given core  (the
measurement of core sizes is described in \S~\ref{sizesec}).
Protostellar sources with bolometric  temperature $T_{bol}<300$~K
(Enoch et al. 2008, in prep.) are indicated by bold symbols.  In general,
the coldest objects, those expected to be embedded within millimeter
cores,  are located within $1.0\times \theta_{1mm}$ of a 1.1~mm core
position.  Thus we define protostellar  cores to be those cores which
have a candidate protostar located within  $1.0 \times \theta_{1mm}$
of the core center.  Note that protostellar cores are defined by any
candidate protostar within $1.0 \times \theta_{1mm}$, not only those
with $T_{bol}<300$~K.

The right panel of Figure~\ref{jdistfig} demonstrates what we would
expect if the same sample of \textit{Spitzer} sources was distributed
randomly over the Bolocam map area.  Although there are approximately
the same number of sources in the random sample as in the candidate
protostar sample, most do not appear on the plot because they are
located much farther than $2 \times \theta_{1mm}$ from the nearest
core.  There are 5 sources from the random sample located within
$1.0\times \theta_{1mm}$ of a core position; thus we can expect a few
false associations between 1.1~mm cores and candidate protostars based
on this criteria.  Indeed, some associations between cores and
\textit{Spitzer} sources are probably just  projections on the sky.
We are especially skeptical of associations where the 1.1~mm core flux
density is much higher than the $70~\micron$ flux.  Using $0.5\times
\theta_{1mm}$ would be a more restrictive choice, but might
mis-identify some protostellar cores as starless.  As we do not want
to contaminate our starless core sample with more evolved sources, we
adopt the more conservative criteria.  Thus the number of starless
cores in each cloud is likely a lower limit to the true value; using
$0.5 \times \theta_{1mm}$ would result in 3 more starless cores in
Perseus, 2 in Serpens, and 4 in Ophiuchus.

\section{Comparing the Starless and Protostellar 1.1~mm Core Populations}\label{embsl}

\begin{figure*}[!ht]
\includegraphics[width=7.3in]{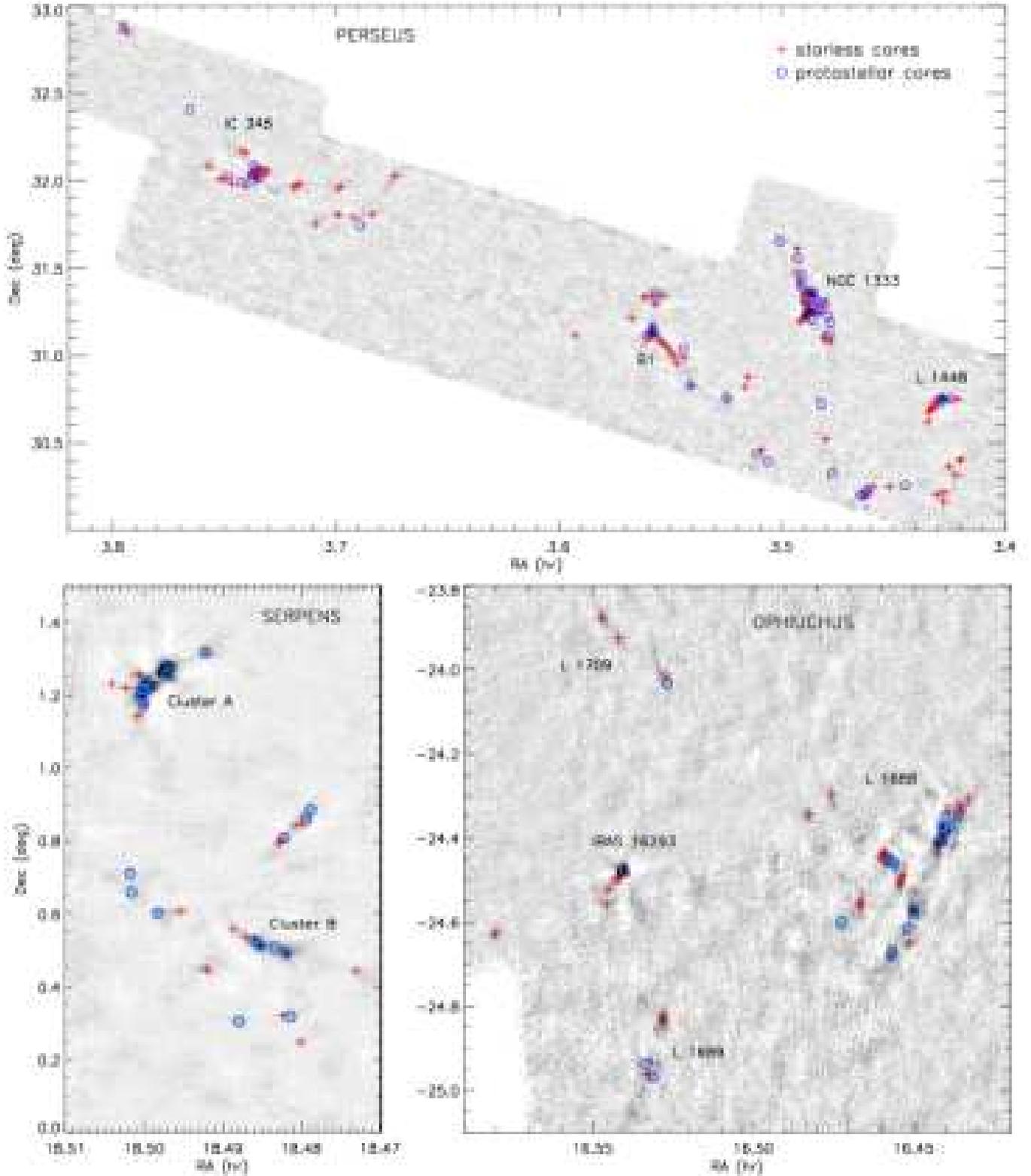}
\caption
{Bolocam maps of Perseus, Serpens, and Ophiuchus, with the positions
  of starless and protostellar cores indicated.  Identified cores have
  peak flux densities of at least $5\sigma$, where $\sigma$ is the
  local rms noise level (on average $\sigma = 15$ mJy beam$^{-1}$ in
  Perseus, 10 mJy beam$^{-1}$ in Serpens, and 25 mJy beam$^{-1}$ in
  Ophiuchus).  Note that due to the large scales, individual
  structures are difficult to see.  Regions of the maps with no
  detected sources have been trimmed for this figure.  Starless and
  protostellar cores cluster together throughout each cloud, with both
  populations tending to congregate along filamentary cloud
  structures.  There are a few exceptional regions, however, which are
  dominated by either starless or protostellar cores (e.g. the B1
  Ridge, Serpens Cluster A).}
\label{spatial}
\end{figure*}

Tables~\ref{sltab} and \ref{pstab} list the Bolocam identifications,
positions, and peak flux densities (from Papers I--III) of starless
and protostellar cores in each cloud.  We find a total of 108 starless
and 92 protostellar cores in the three cloud sample.  Cores are
identified based on a peak flux density at least 5 times  the local
rms noise level, and positions are determined using a surface
brightness-weighted centroid, as described in Paper III.  The Bolocam
maps of Perseus, Serpens, and Ophiuchus are shown in
Figure~\ref{spatial}, with the positions of starless and protostellar
cores indicated.  The average $1\sigma$ rms is 15~mJy beam$^{-1}$ in
Perseus, 10~mJy beam$^{-1}$ in Serpens, and 25 mJy beam$^{-1}$ in
Ophiuchus.  

\begin{deluxetable*}{lccccc}
\tablecolumns{6}
\tablewidth{0pc}
\tablecaption{Statistics of 1.1~mm cores in the three clouds\label{coretab}}
\tablehead{
\colhead{Cloud} & \colhead{$\mathrm{N_{total}}$\tablenotemark{1}\ \ } & 
\colhead{\ \ \ $\mathrm{N_{\mathrm{SL}}}$\tablenotemark{2}\ \ } & 
\colhead{\ \ \ $\mathrm{N_{\mathrm{PS}}}$\tablenotemark{3} \ \ } & 
\colhead{ \ \ $\mathrm{N_{\mathrm{SL}}/N_{\mathrm{PS}}}$ \ } & 
\colhead{$\mathrm{N_{\mathrm{PS}}}$ (mult)\tablenotemark{4}}
}
\startdata
Perseus   &  122 &  67  &  55  &  1.2  &   13 \\ 
Serpens   &  35  &  15  &  20  &  0.8  &   11 \\ 
Ophiuchus &  43  &  26  &  17  &  1.5  &   3  
\enddata
\tablenotetext{1}{Total number of identified 1.1~mm cores.}
\tablenotetext{2}{Number of starless 1.1~mm cores, i.e., cores that do not have 
a protostar candidate located within $1.0\times
\theta_{1mm}$ of the core position.}
\tablenotetext{3}{Number of protostellar cores.}
\tablenotetext{4}{Number of protostellar cores that are associated with more than 
one candidate cold protostar (each within $1.0\times \theta_{1mm}$ of the core position).}
\end{deluxetable*}

General core statistics, including  the number of starless ($N_{\mathrm{SL}}$)
and protostellar ($N_{\mathrm{PS}}$) cores in each cloud, as well as the ratio
$N_{\mathrm{SL}}/N_{\mathrm{PS}}$, are given in Table~\ref{coretab}.  Note that the
number of starless and protostellar cores are approximately equal in
each cloud ($N_{\mathrm{SL}}/N_{\mathrm{PS}}=1.2$  in Perseus, 0.8 in Serpens, and 1.4
in Ophiuchus), a fact that will be important for our discussion of the
starless core lifetime  in \S\ref{lifesec}.   The last column of
Table~\ref{coretab} gives the number of individual 1.1~mm cores that
are associated  with more than one candidate protostar (each located
within $1 \times \theta_{1mm}$ of the core position).  There are 13
such ``multiple'' protostellar sources in Perseus (24\% of the
protostellar core sample), 11 in Serpens (55\%), and 3 in Ophiuchus
(17\%).  In general there are two or three candidate protostars
associated with each ``multiple'' core, with the exception of one core
in Serpens (5 candidate protostars).  Throughout this work multiple
protostellar cores are treated as single objects. 

We now compare the physical properties of the starless and protostellar 
core populations in each cloud, with two primary
goals.  Isolating a starless sample allows us to probe the
initial conditions of star formation, and differences between the
starless and protostellar core samples are indicative of how the
formation of a central protostar alters core properties.  In the 
following sections we follow the methodology of Paper III,
examining the sizes and shapes of cores, their peak and mean
densities, distributions of core mass versus size, spatial
clustering properties, and relationship to the surrounding cloud
column density.

\subsection{Sizes and Shapes}\label{sizesec}

\begin{figure*}
\vspace{-0.8in}
\centering
\includegraphics[angle=90,width=6.5in]{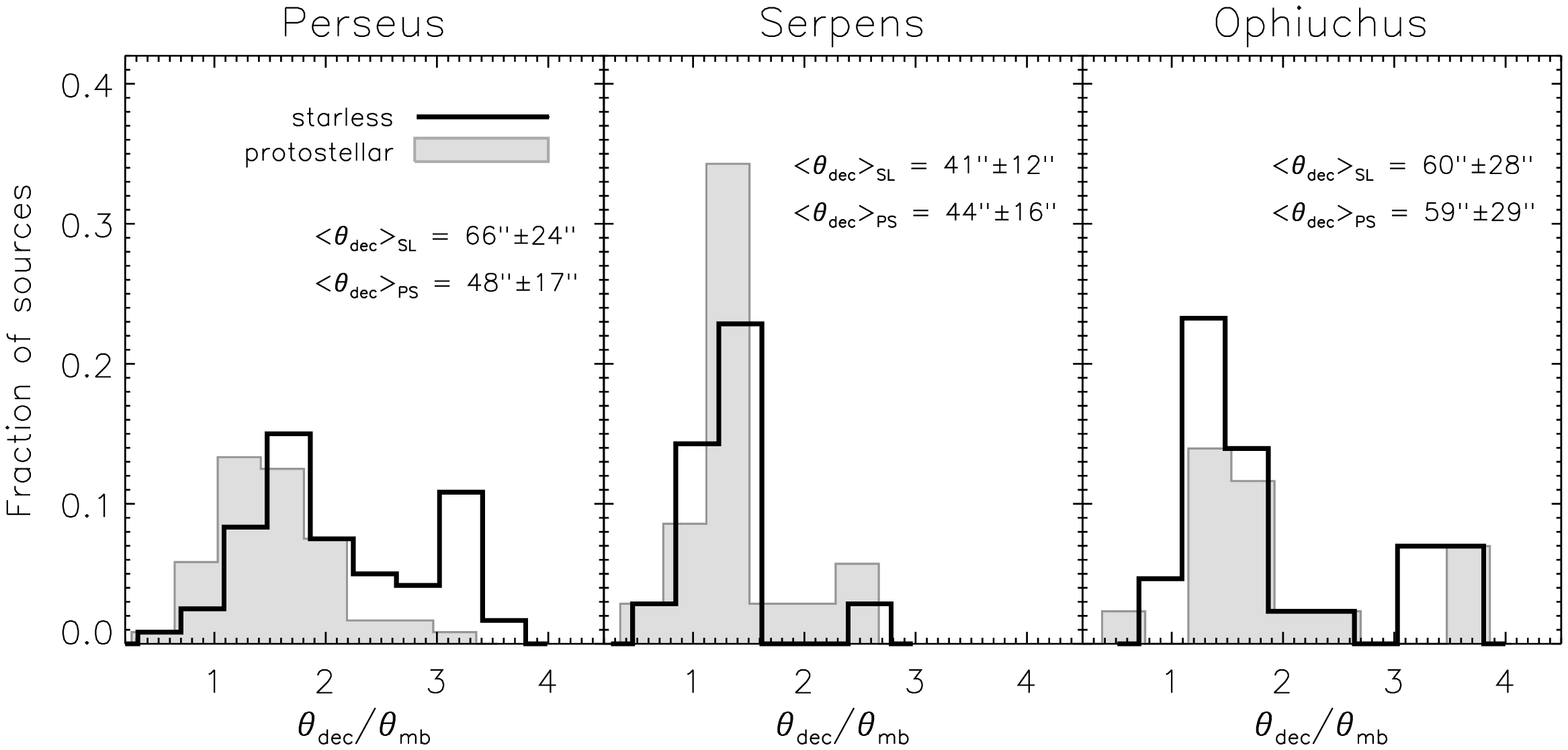}
\vspace{-0.7in}
\caption
{Distributions of the angular deconvolved sizes of starless and
  protostellar cores in the three clouds.  The size is given in units
  of the beam FWHM ($\theta_{mb}$), and the mean of each distribution
  $\pm$  the dispersion in the sample is listed.  Starless cores are
  larger on average than protostellar cores in Perseus, with a
  flattened distribution out to $3.5~ \theta_{mb}$.  In Serpens and
  Ophiuchus, however, there is little difference between the starless
  and protostellar distributions, and there are fewer very large
  cores, particularly in Serpens.  The value of
  $\theta_{dec}/\theta_{mb}$ can be used to infer the steepness of the
  source radial density profile (see text and \citealt{young03}).}
\label{sizecomp}
\end{figure*}

Source angular FWHM sizes ($\theta_{meas}$) are measured by fitting an
elliptical Gaussian after masking out nearby sources using a mask
radius equal to half the distance to the nearest neighbor (see Paper
II).  The angular deconvolved core size is the geometric mean of the
deconvolved minor and major angular FWHM sizes: $\theta_{dec} = \sqrt{
  \theta_{d,maj} ~  \theta_{d,min}}$, where  $\theta_{d} = \sqrt{
  \theta_{meas}^2 -  \theta_{mb}^2}$ and $\theta_{mb}=31\arcsec$ is
the beam FWHM.  Deconvolved sizes for starless and protostellar cores
are given in Tables~\ref{sltab} and \ref{pstab},  respectively, and
the size distributions are plotted in Figure~\ref{sizecomp}.  

As discussed in Paper III, the measured size does not necessarily
represent a physical boundary, but rather is a characteristic scale
that depends on the linear resolution and intrinsic source density
profile.   For sources with power law density profiles, which do not
have a well defined size, $\theta_{dec}/\theta_{mb}$ is independent of
distance and simply related to the index of the power law
\citep{young03}.  According to the correlation between
$\theta_{dec}/\theta_{mb}$ and density power law exponent $p$ found by
\citet{young03}, a mean $\theta_{dec}/\theta_{mb}$ value of
$50\arcsec/31\arcsec = 1.6$ for protostellar cores in Perseus implies
an average power law index of $p\sim1.4$ to 1.5.  Many well-known
protostellar sources have been found to have envelopes consistent with
power law density profiles, as determined by  high-resolution imaging
combined with radiative transfer modeling.  Our inferred average index
for Perseus protostellar cores ($p=1.4-1.5$) is consistent with the
mean $p\sim1.6$ from radiative transfer modeling of Class~0 and
Class~I envelopes \citep{shir02,young03}.  For reference, a singular
isothermal sphere (SIS) has $p=2$ ($\theta_{dec}/\theta_{mb} \sim
0.9$),  and the profile  expected for a free-falling envelope is
$p=1.5$ ($\theta_{dec}/\theta_{mb} \sim 1.6$) \citep{shu77,young03}.

Starless cores in Perseus are larger on average than protostellar
cores (Figure~\ref{sizecomp}); the starless distribution  is
relatively flat, with a few barely resolved cores, and several larger
than 3~$\theta_{mb}$.  The Student's T-test confirms that the mean 
sizes of starless and protostellar cores are significantly different 
(significance of the T statistic is $2\times 10^{-5}$).
A mean $\theta_{dec}/\theta_{mb}$ of 2.2 for
starless cores in Perseus would imply an extremely shallow mean power
law index of $p\sim1.1$, and the maximum value
($\theta_{dec}/\theta_{mb}\sim3.5$) would correspond to $p<0.8$.
Other flattened profiles, such as the Bonnor-Ebert (BE) sphere
\citep{ebert55,bonn56}, could also produce large
$\theta_{dec}/\theta_{mb}$ values.   A BE profile with a central
density of $10^5$~cm$^{-3}$ and an outer radius of $6\times10^4$~AU
would correspond to $\theta_{dec}/\theta_{mb}=2.0$ at the distance of
Perseus.  There is significant observational evidence that many
starless cores do indeed look like BE spheres (e.g.,
\citealt{john00,shir00,alv01,evans01}).   We conclude that very large starless
cores are more consistent with BE spheres or other flattened density
profiles ($p\lesssim 1$) than with the classical pre-collapse SIS
($p=2$).
\footnote{As the pressure-truncated boundary radius of the BE
  solution has not been observationally verified, the resemblance
  between starless cores and BE profiles is necessarily limited to the
  region inside the boundary radius.  Also note that a sufficiently
  centrally condensed BE sphere will be indistinguishable from a power
  law on the scales to which we are sensitive.}

\begin{figure*}
\vspace{-0.8in}
\centering
\includegraphics[angle=90,width=6.5in]{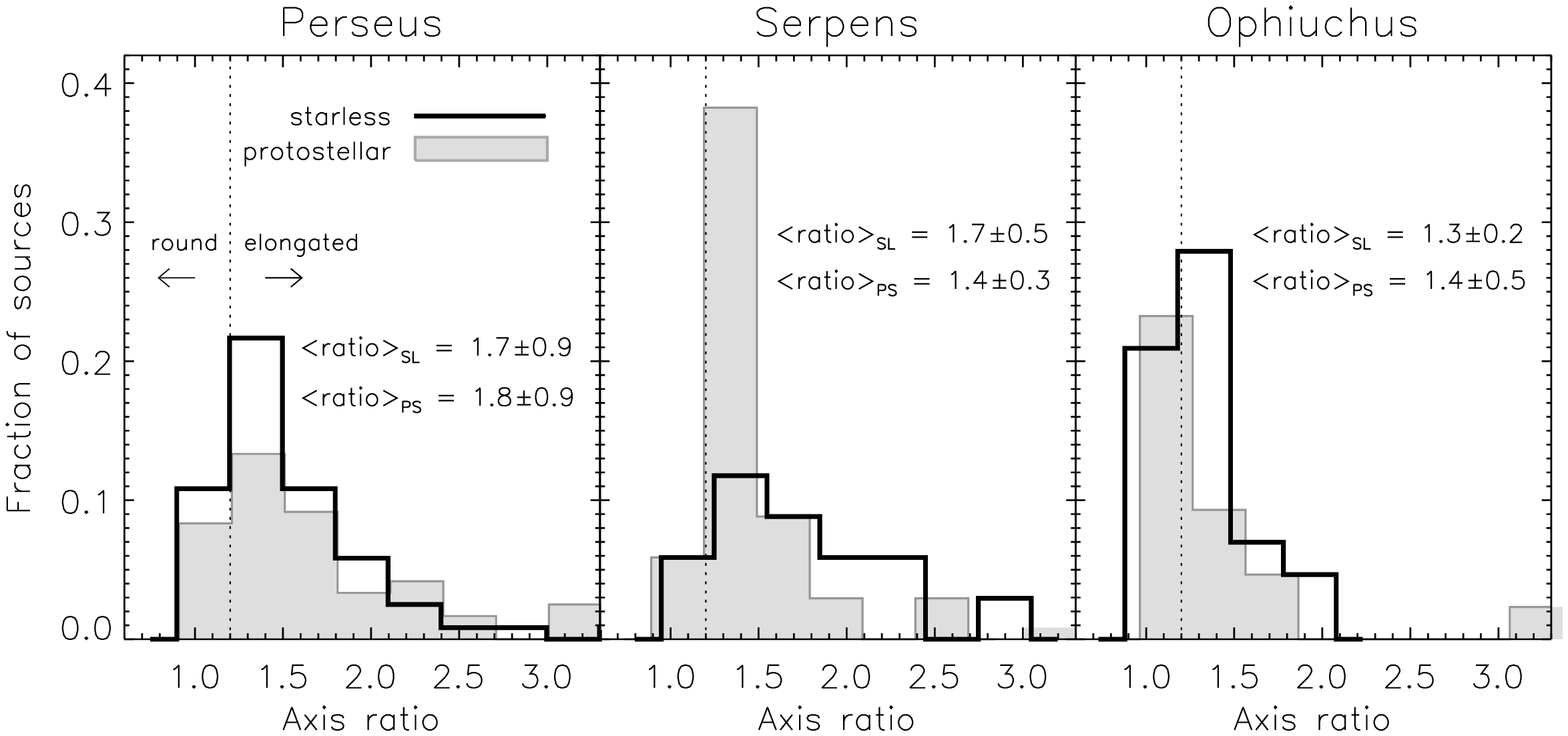}
\vspace{-0.7in}
\caption
{Distributions of the axis ratios of starless and protostellar cores.
  Cores with axis ratios $<1.2$ are considered round and those with
  axis ratios $>1.2$ elongated, based on Monte Carlo simulations
  (Paper I).  Only in Serpens, where starless cores tend to be
  slightly more elongated than protostellar cores, is there a
  distinguishable difference between the starless and protostellar
  populations.  }\label{axiscomp}
\end{figure*}

Despite their size differences, there is little difference in the axis
ratios of starless and protostellar cores in Perseus
(Figure~\ref{axiscomp}).  Both populations are slightly elongated on
average.  The axis ratio  is defined at the half-max contour, using
deconvolved sizes: $\theta_{d,maj} / \theta_{d,min}$.  Values for
individual cores are given in Tables~\ref{sltab} and
\ref{pstab}. Starless and protostellar cores have mean
axis ratios of $1.7 \pm 0.9$ and $1.8 \pm 0.9$.   Standard deviations
quoted here and in the rest of \S\ref{embsl} are dispersions in the
sample, \textit{not} errors in the mean.  Monte Carlo tests (Paper I)
indicate that cores with axis ratios less than 1.2 should be
considered round.  
 
In contrast to Perseus, starless cores in Serpens are no larger than
protostellar cores.  Mean values for both ($\theta_{dec}/\theta_{mb}
\sim 1.5$) correspond to an average power law index of $p=1.5$,
similar to that found for protostellar  cores in Perseus, and to
typical radiative transfer modeling results for  protostellar
envelopes.  Starless cores in Serpens may be more elongated than
protostellar cores (mean axis ratios of  $1.7 \pm 0.5$ and
$1.4\pm0.3$, respectively), but the difference is not statistically
significant.

In Ophiuchus there is no measurable difference between the starless
and protostellar populations.  The starless and protostellar samples
have similar mean sizes, and both  display a bimodal behavior
(Figure~\ref{sizecomp}).  The lower peak is similar to the single peak
seen in Serpens at $\theta_{dec} = 1 - 2~\theta_{mb}$,  while the
smaller upper peak is at sizes of $\theta_{dec} = 3 - 4~ \theta_{mb}$,
comparable to the largest starless cores in Perseus.  The mean
$\theta_{dec}/\theta_{mb}$ for protostellar cores (1.9) corresponds to
an average power law index of $p=1.3$.  
Both starless and protostellar
cores in Ophiuchus  appear fairly round, with the mean axis ratios of
$1.3\pm0.2$ and $1.4\pm0.5$, respectively (Figure~\ref{axiscomp}).  As
discussed in Paper III, larger axis ratios in Perseus and Serpens may
be at least partly an effect of the lower linear resolution in those
clouds compared to Ophiuchus (e.g. blending).

Were we to calculate linear sizes, cores in Ophiuchus would appear
smaller by nearly a factor of two compared to Perseus and Serpens,
given the smaller distance to Ophiuchus. This is primarily a
systematic effect of the linear resolution, however.  In Paper~III we
found that convolving the Ophiuchus map with a larger beam to match
the linear resolution of Perseus and Serpens produced larger linear
core sizes by nearly a factor of two, but similar measured
$\theta_{dec}/\theta_{mb}$ values.  Thus we focus here on angular
sizes only.

To summarize, protostellar cores in all three clouds have mean sizes
consistent with power law density profiles with an average index
$p=1.3-1.5$.  Starless cores in Perseus are significantly larger on
average than protostellar cores, suggestive of BE spheres or
other shallow density profiles.   Starless cores are quite compact in
Serpens, while both starless and protostellar cores in Ophiuchus
display a bimodal distribution of sizes, with a few very large cores.
The deficit of cores with $\theta_{dec}/\theta_{mb} > 2$ in Serpens
and Ophiuchus as compared to Perseus may be related to the general lack
of isolated sources in those clouds; the measured size of a core is
limited by the distance to the nearest neighboring source, so sizes
will tend to be smaller in crowded regions (e.g. L~1688 Ophiuchus) 
than for isolated sources.

\subsection{Core Masses and Densities}\label{densec}

Core total masses, given in Tables~\ref{sltab} and \ref{pstab}, are
calculated from the total 1.1~mm flux, $S_{1.1mm}$:
\begin{equation}
M = \frac{d^2 S_{\nu}}{B_{\nu}(T_D) \kappa_{\nu}}, \label{masseq}
\end{equation}
where $d$ is the cloud distance, $B_{\nu}$ is the Planck function at
dust temperature $T_D$, and $\kappa_{1.1mm} = 0.0114$ cm$^2$ g$^{-1}$
is the dust opacity.  The total 1.1~mm flux density
is integrated in the largest aperture ($30\arcsec-120\arcsec$
diameters in steps of $10\arcsec$) that is smaller than the distance
to the nearest neighboring source.  We assume that the dust
emission at $\lambda=1.1$~mm is optically  thin, and that $T_D$ and
$\kappa_{1.1mm}$ are independent of position within a core.  The value
of $\kappa_{1.1mm}$ is interpolated from Table~1 column~5 of
\citet{oh94} for dust grains with thin ice mantles, and includes a gas
to dust mass ratio of 100.  A recent measurement of $\kappa$ based on
near-IR data and $450 \micron$ and $850 \micron$ SCUBA maps yields
$\kappa_{1.1mm} = 0.0088$ cm$^2$ g$^{-1}$ \citep{shir07}, which would
increase our masses by a factor of 1.3.

The value of $T_D$ should depend on whether a core is starless
or has an internal source of luminosity, so we assume a slightly
higher temperature for protostellar cores ($T_D=15$~K) than for
starless cores ($T_D=10$~K).   For dense regions without internal
heating, the mean temperature is about 10~K, warmer on the outside and
colder on the inside \citep{evans01}.  A recent NH$_3$ survey of the 
Bolocam cores in Perseus confirms that the median kinetic temperature 
of starless cores in Perseus is 10.8~K (see Schnee et al. 2008, in prep, 
and \S~\ref{boundsec}).  Our assumed value of 15~K for
protostellar cores is the average isothermal dust temperature found
from radiative transfer models of Class 0 and Class I
protostars \citep{shir02,young03}.  The isothermal dust  temperature
is the temperature that, when used in an isothermal mass equation
(e.g., Eq.~\ref{masseq}) yields the same mass as a detailed
radiative transfer model including temperature gradients.   
There is a factor of 1.9 difference in mass between assuming 
$T_D=10$~K and 15~K.

\begin{figure*}
\centering
\vspace{-0.7in}
\includegraphics[angle=90,width=6.5in]{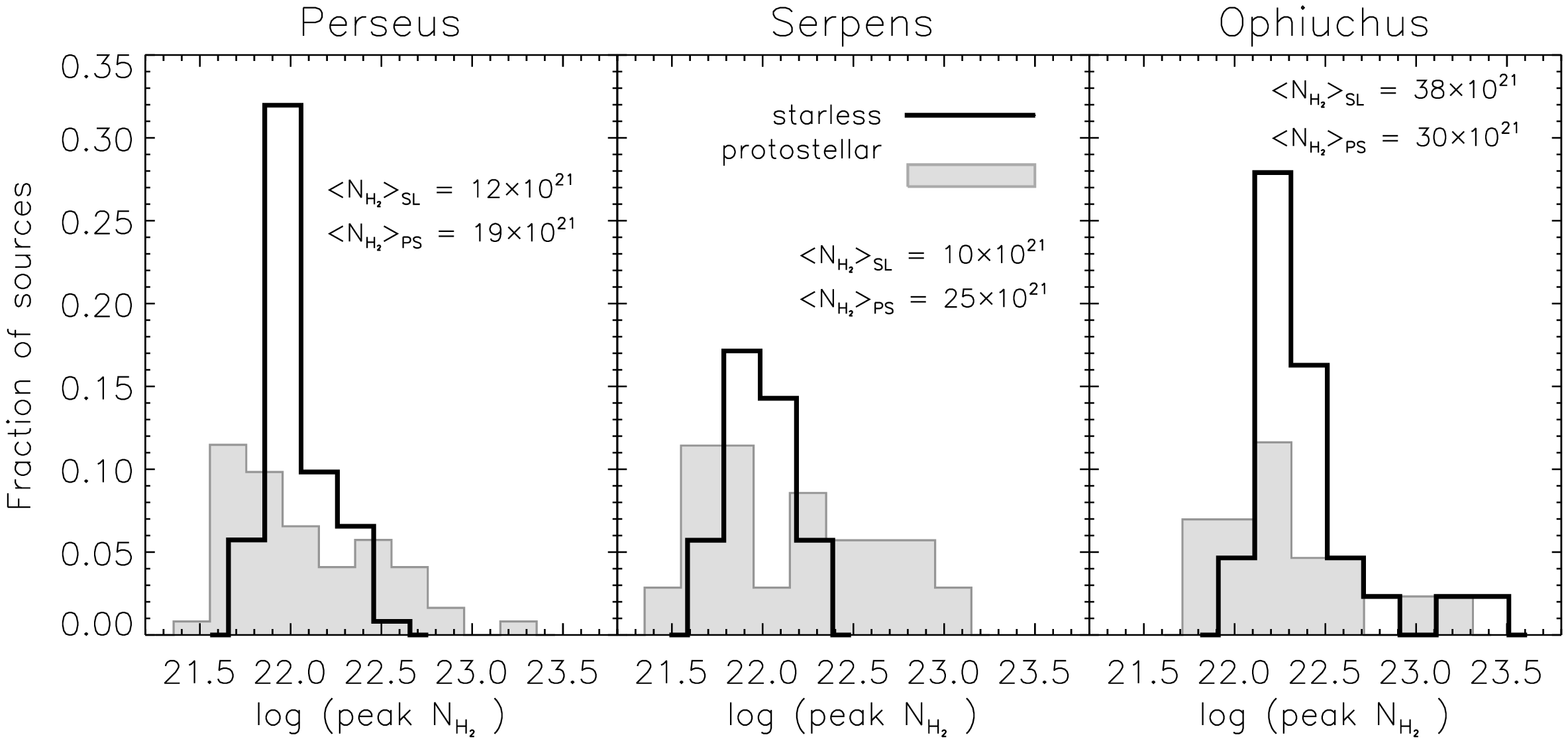}
\vspace{-0.7in}
\caption
{Distributions of the peak column density \nh\ of starless and
  protostellar cores.  The peak \nh\ values of starless cores are
  considerably lower than those of protostellar cores in both Perseus
  and Serpens, by factors of approximately 1.6 and 2.5, respectively.
  In Ophiuchus there is no significant difference in the mean values,
  but in all three clouds the starless distribution is confined to a 
  narrowly peaked distribution.   
}\label{pdencomp}
\end{figure*}

\begin{figure*}
\centering
\vspace{-0.7in}
\includegraphics[angle=90,width=6.5in]{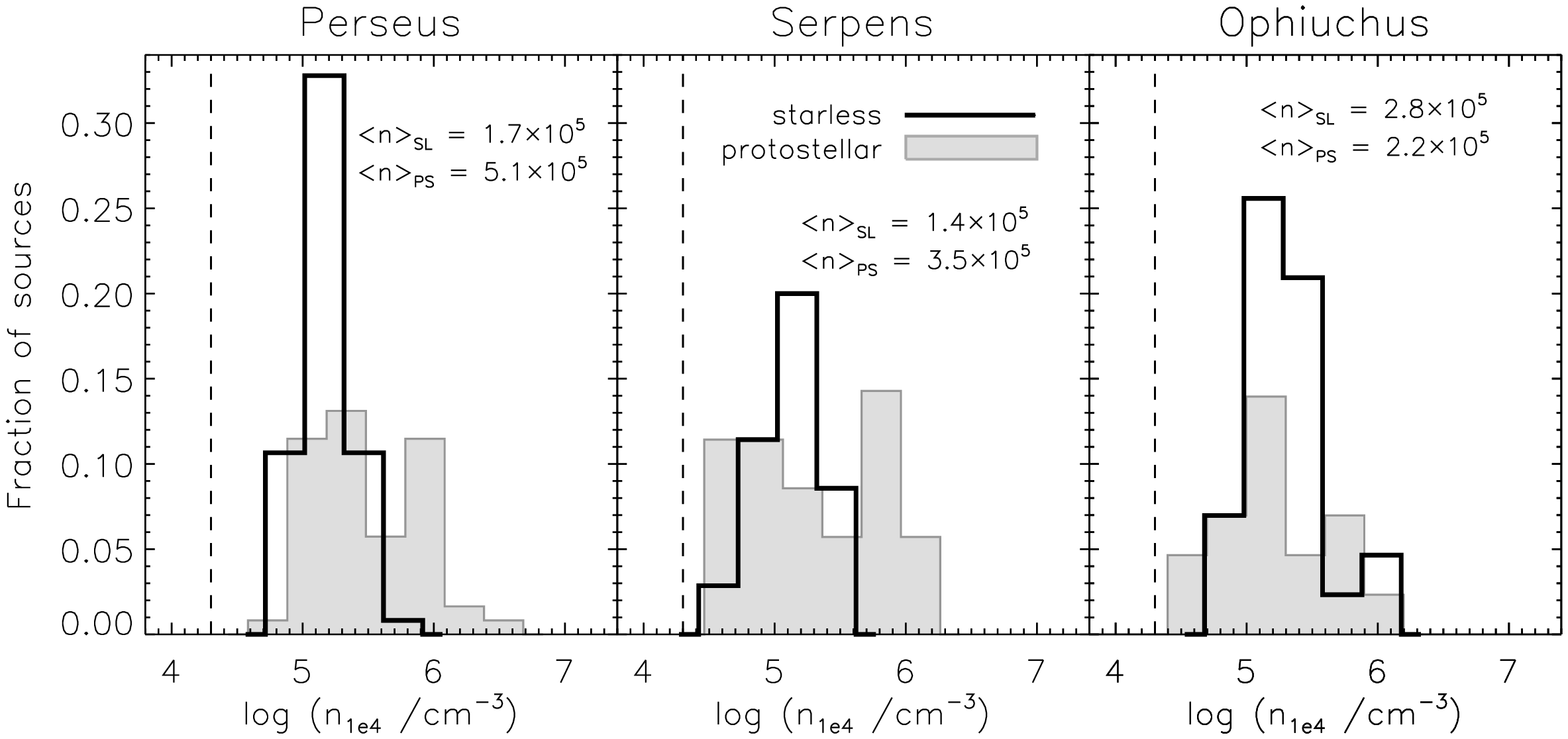}
\vspace{-0.7in}
\caption
{Distributions of the mean density $n_{1e4}$ of starless and
  protostellar cores, where the density is calculated in a fixed
  linear aperture of diameter $10^4$~AU (\S~\ref{densec}).  As for
  the peak \nh\ distribution (Figure~\ref{pdencomp}), starless cores
  tend to have lower mean densities than protostellar cores in Perseus
  and Serpens, by approximately a factor of three.  In Ophiuchus, by
  contrast, there is almost no difference between the two
  distributions.  Note that the Bolocam surveys are only sensitive to
  relatively dense cores with $n \gtrsim 2\times 10^4$~cm$^{-3}$
  (dashed lines, see Paper III).}
\label{mdencomp}
\end{figure*}

Figures~\ref{pdencomp} and \ref{mdencomp} compare the peak and mean
densities of starless and protostellar cores in each cloud.  We use
the peak column density \nh, calculated from the peak 1.1~mm flux
density $S_{1.1mm}^{beam}$, as a measure of the peak density:  
\begin{equation}
N(\mathrm{H}_2) = \frac{S_{\nu}^{beam}}{\Omega_{beam} \mu_{H_2} m_H
  \kappa_{\nu} B_{\nu}(T_D)}, \label{aveq}
\end{equation}
with $N($H$_2)/A_V = 0.94 \times 10^{21}$ cm$^{-2}$~mag$^{-1}$
\citep{flw82}.  Here $\Omega_{beam}$ is the beam solid angle, $m_H$ is
the mass of hydrogen, and $\mu_{H_2}=2.8$ is the mean molecular weight
per H$_2$ molecule.   As discussed in Paper III, the 1.1~mm emission
detected by Bolocam traces significantly higher column densities than
other tracers such as the reddening of background stars.  

Mean particle densities, given in Tables~\ref{sltab} and \ref{pstab},
are calculated within a fixed linear aperture of diameter $10^4$~AU: 
\begin{equation}
n_{1e4} = \frac{3M_{1e4}}{4 \pi (D_{1e4}/2)^3 \mu m_H}, \label{deneq}
\end{equation}
where $D_{1e4}$ is the aperture size ($10^4$~AU),  $M_{1e4}$ is the
mass calculated from the 1.1~mm flux within that aperture, and
$\mu=2.33$ is the mean molecular weight per particle.  An aperture of
$10^4$~AU corresponds to approximately $40\arcsec$ in Perseus and
Serpens, and $80\arcsec$ in Ophiuchus.   Note that $n_{1e4}$ does not 
depend on the core size; while in some cases
$10^4$~AU may be considerably smaller or larger than  the source FWHM,
a fixed linear aperture is used here to mitigate the effects  of
linear resolution, which was found in Paper III to significantly bias
the mean density calculated within the FWHM contour.

The average peak column density (\nh) for starless cores in Perseus is
$12\times10^{21}$~cm$^{-2}$, while the average for protostellar cores
is 50\% higher, $19\times10^{21}$~cm$^{-2}$, with a much wider
dispersion (the significance of the Student's T-statistic is 0.03).
Similarly, the typical mean density ($n_{1e4}$) of starless cores
($1.7\times10^5$~cm$^{-3}$) is a factor of three smaller than that of
protostellar cores ($5.1\times10^5$~cm$^{-3}$; Student's T-test
significance $6\times10^{-5}$).  The large difference in mean
densities is due primarily to the significantly smaller sizes of
protostellar cores in Perseus.  Recently, \citet{jorg07} found a
similar result for Perseus, comparing SCUBA $850~ \micron$ cores with
and without internal luminosity sources, as determined using
\textit{Spitzer} c2d data. Those authors concluded that cores with
embedded YSOs (located within 15\arcsec\ of the core position) have
higher ``concentrations'' on average.  Note that average $n_{1e4}$
values are much larger than the minimum detectable density of $\sim 2
\times 10^4$ cm$^{-3}$ from Paper III, due in part to few sources near
the detection limit and in part to the difference in calculating
densities in a fixed aperture.

Density distributions  for starless and protostellar cores in Serpens
are similar to those in Perseus.  Peak \nh\ values are substantially
smaller for starless ($\mean{\nh}=10\times10^{21}$~cm$^{-2}$) than for
protostellar ($\mean{\nh}=25\times10^{21}$~cm$^{-2}$; Student's T-test
significance $0.04$) cores, and form a much narrower distribution.
Likewise, typical mean densities of starless cores
($1.4\times10^5$~cm$^{-3}$) are nearly three times smaller than those
of protostellar cores ($3.5\times10^5$~cm$^{-3}$; Student's T-test
significance $0.03$).  In contrast to Perseus, however, mean density
differences in Serpens are due entirely to the higher masses of
protostellar cores, as starless and protostellar cores have similar
sizes in Serpens.  As was the case for core sizes and shapes, there is
essentially no difference between the peak or mean densities of
starless and protostellar cores in Ophiuchus.  Average peak
\nh\ values are similar ($30-40\times10^{21}$~cm$^{-2}$) and the two
distributions have similar dispersions, and average $n_{1e4}$ values
are $2.2 - 2.8\times10^5$~cm$^{-3}$.

To summarize, peak column densities and mean densities of starless
cores in Perseus and Serpens are lower on average than for
protostellar cores, whereas in Ophiuchus there is no significant
difference between the starless and protostellar populations.  The
\nh\ distributions for starless cores are quite narrow in all three
clouds, indicating a  small range of column densities.  This narrow
distribution more likely represents an upper limit to \nh\ for
starless cores, as we are not sensitive to very large low mass cores
(see \S~\ref{mvssec}), but in general cores with the highest column
densities tend to be protostellar.

\begin{figure*}[!hb]
\centering
\includegraphics[width=7.1in]{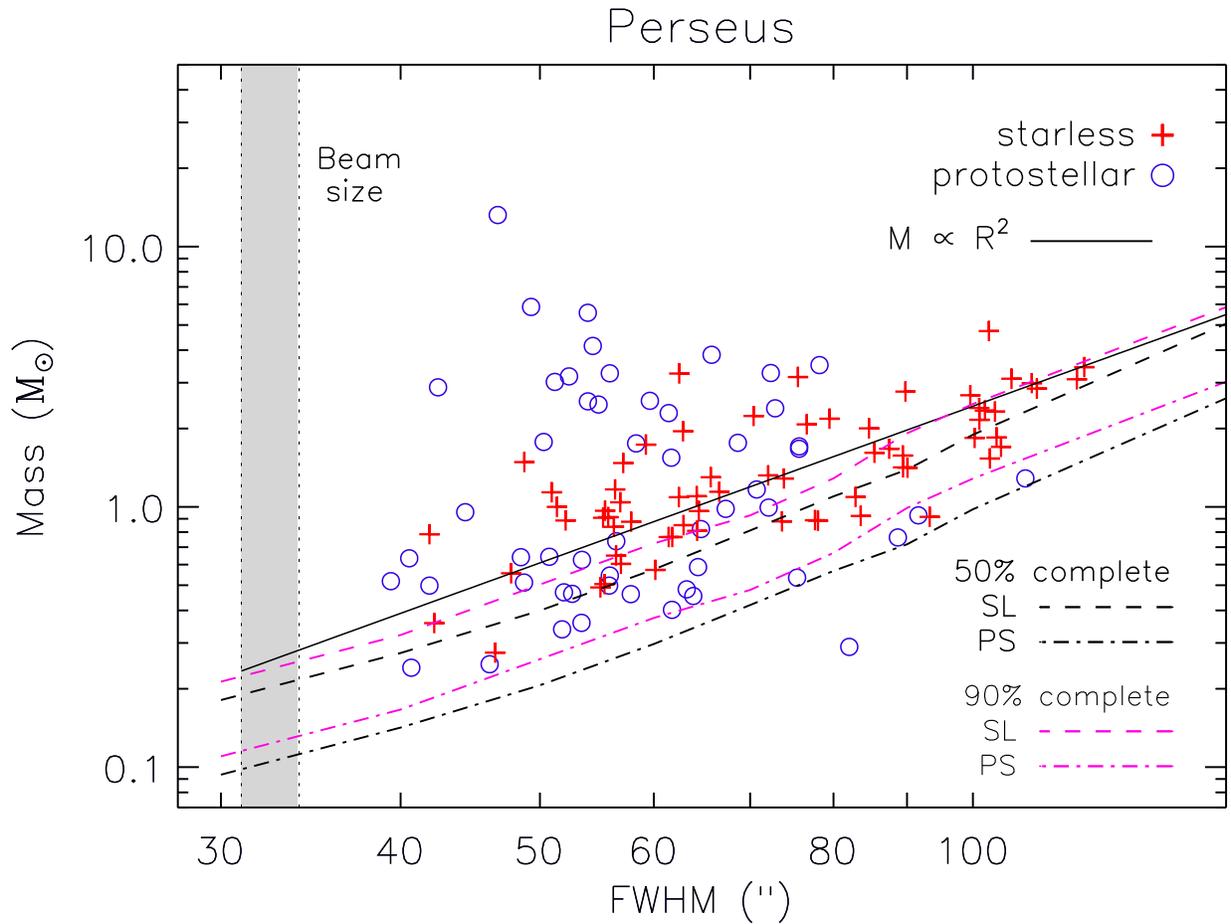}
\caption
{Total mass versus angular FWHM size for starless and protostellar
  cores in Perseus.   Dashed and dash-dot lines indicate empirically
  derived 50\% (and 90\%; light gray) completeness limits for starless
  and protostellar cores, respectively.  The two populations seem to
  inhabit different regions of the parameter space: starless cores
  tend to follow a constant surface density relationship ($M\propto
  R^2$, solid line), consistent with their narrow distribution in
  column density (Figure~\ref{pdencomp}), while protostellar cores
  have a wide range in masses for a relatively small range in sizes.
  This relationship suggests a simple explanation for how the
  protostellar cores might have evolved from a similar population of
  starless cores, with individual cores becoming smaller and denser at
  a constant mass until protostar formation is triggered.}
\label{mvscomp}
\end{figure*}

\begin{figure*}
\centering
\includegraphics[angle=90,width=7.3in]{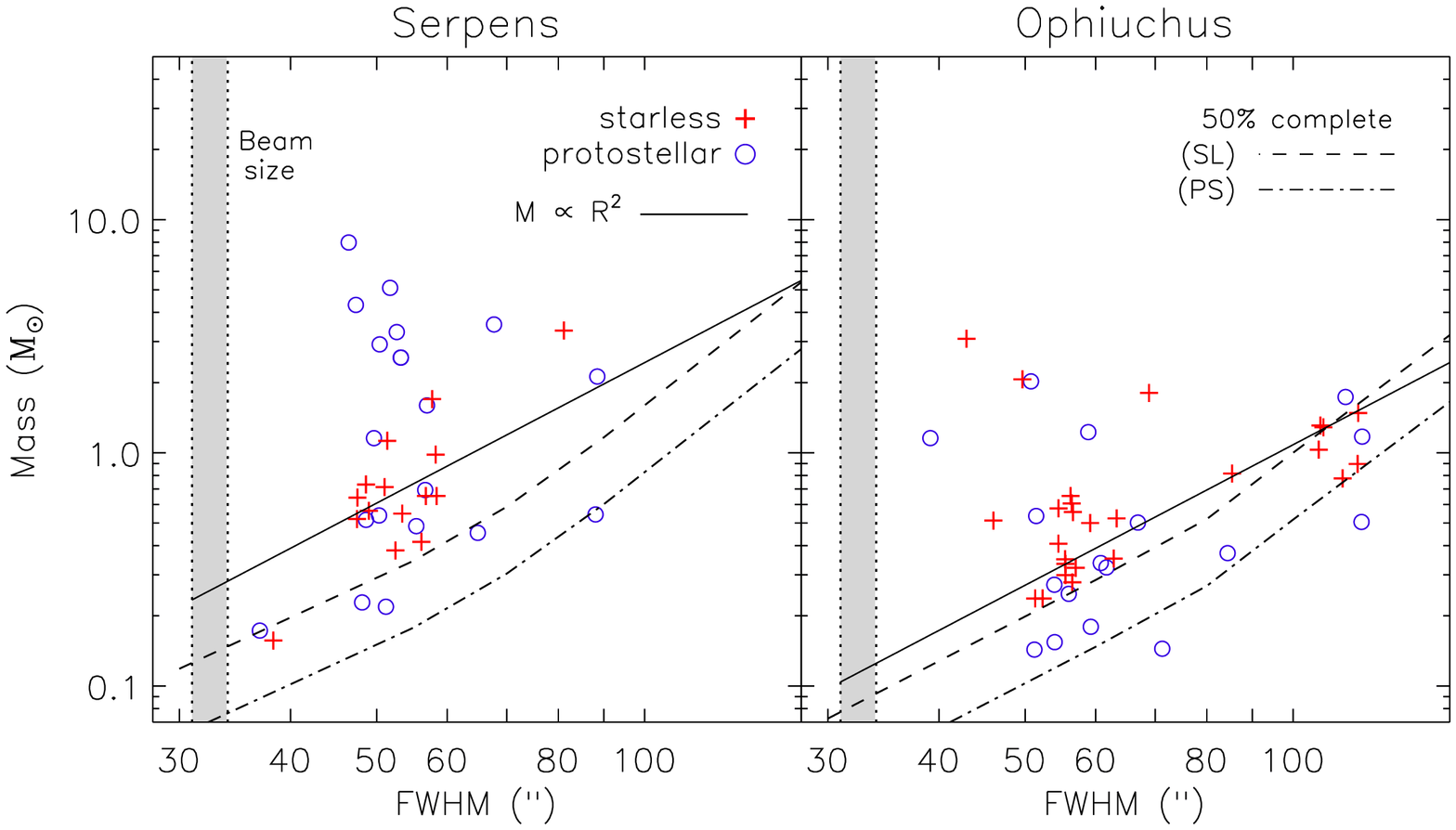}
\vspace{-0.5in}
\caption
{Total mass versus angular FWHM size for starless and protostellar
  cores in Serpens  and Ophiuchus.  Lines are as in
  Figure~\ref{mvscomp}; as for Perseus, the 90\% completeness curves
  are approximately a factor of 1.2 higher in mass than the 50\%
  completeness curves, although they are omitted here for clarity.
  Unlike in Perseus, starless cores do not necessarily follow a
  constant surface density ($M\propto R^2$) line.  Strikingly,  there
  is no population of large starless cores in Serpens; it is not clear
  how the current population of relatively massive protostellar cores
  could have evolved from such compact, low mass starless cores.  This
  discrepancy suggests that future star formation in the cloud
  may result in stars of considerably lower mass than the currently
  forming protostars.}  
\label{mvscomp2}
\end{figure*}

\subsection{The Mass versus Size Distribution}\label{mvssec}

Figure~\ref{mvscomp} plots total core mass versus angular FWHM
size for starless and protostellar cores in Perseus.   The beam size
and empirical 50\% and 90\% completeness limits as a function of size, derived
from Monte Carlo simulations (Paper I), are indicated.  Completeness
limits are lower for protostellar cores by a factor of two due to
the higher dust temperature assumed in the mass calculation (15~K) 
compared to starless cores (10~K).  There is approximately a factor 
of 1.2 increase in mass between the 50\% and 90\% completeness curves, 
which holds true for the other clouds as well. 

In this diagram, starless cores seem to follow a
constant surface density ($M \propto R^2$) curve, consistent
with the narrow distribution of peak column densities
(Figure~\ref{pdencomp}).   Protostellar cores, in contrast, have a
narrower range of sizes for a somewhat larger range of masses.   This
is a restatement of the results from \S~\ref{sizesec} and
\S~\ref{densec}: protostellar cores in Perseus are smaller and have
higher mean densities than starless cores.  Note that because our
completeness limits are similar to $M \propto R^2$, the distribution
of starless cores in mass and size only implies that the upper
envelope of cores follows a constant surface density curve.  For
example, there could be a population of large, low mass cores that we
are unable to detect.  Nevertheless, the two populations seem to fill
different regions of the mass versus size parameter space. 

Examining Figure~\ref{mvscomp}, it is easy to imagine how protostellar
cores in Perseus might have evolved from the current population of
starless cores, by decreasing in size and increasing in density for a
constant mass, until collapse and protostellar formation is triggered.
Equivalently, the formation of a central protostar within a previously
starless core is associated with a decrease in core size and an
increase in core density.  

Such a simple scenario is not consistent with the other clouds,
however, as is evident in Figure~\ref{mvscomp2}.  Although
protostellar cores in Serpens have a small range in sizes for a large
range of masses, as seen in Perseus, there is no population of large
starless cores in Serpens.  In fact, it is unclear how the relatively
massive protostellar cores ($\mean{M}_{\mathrm{PS}}=2.1$) in Serpens could have
evolved from the current population of compact,  low mass starless
cores ($\mean{M}_{\mathrm{SL}}=0.9$).  It appears that Serpens has exhausted
its reserve of starless cores with $M \gtrsim 2 \msun$; unless new
cores are formed from the lower density medium, future star
formation in the cloud will result in stars of considerably lower mass
than the current protostellar population.

This deficit of starless cores at higher core masses  may be the
result of a mass dependence in the timescale for protostellar
formation, with higher mass cores forming protostars more quickly
\citep{hatch08}.  In both Serpens and Perseus the most massive cores
tend to be protostellar in nature, although the trend is much more
extreme in Serpens.   In Ophiuchus, where core masses are lower on
average than in the other clouds, no such distinction between starless
and protostellar cores is seen.  There is essentially no difference
between  the starless and protostellar populations in Ophiuchus,
suggesting very little core evolution  after the formation of a
protostar.

\subsection{Relationship to Cloud Column Density}\label{probsec}

\begin{figure*}
\includegraphics[angle=90,width=7.3in]{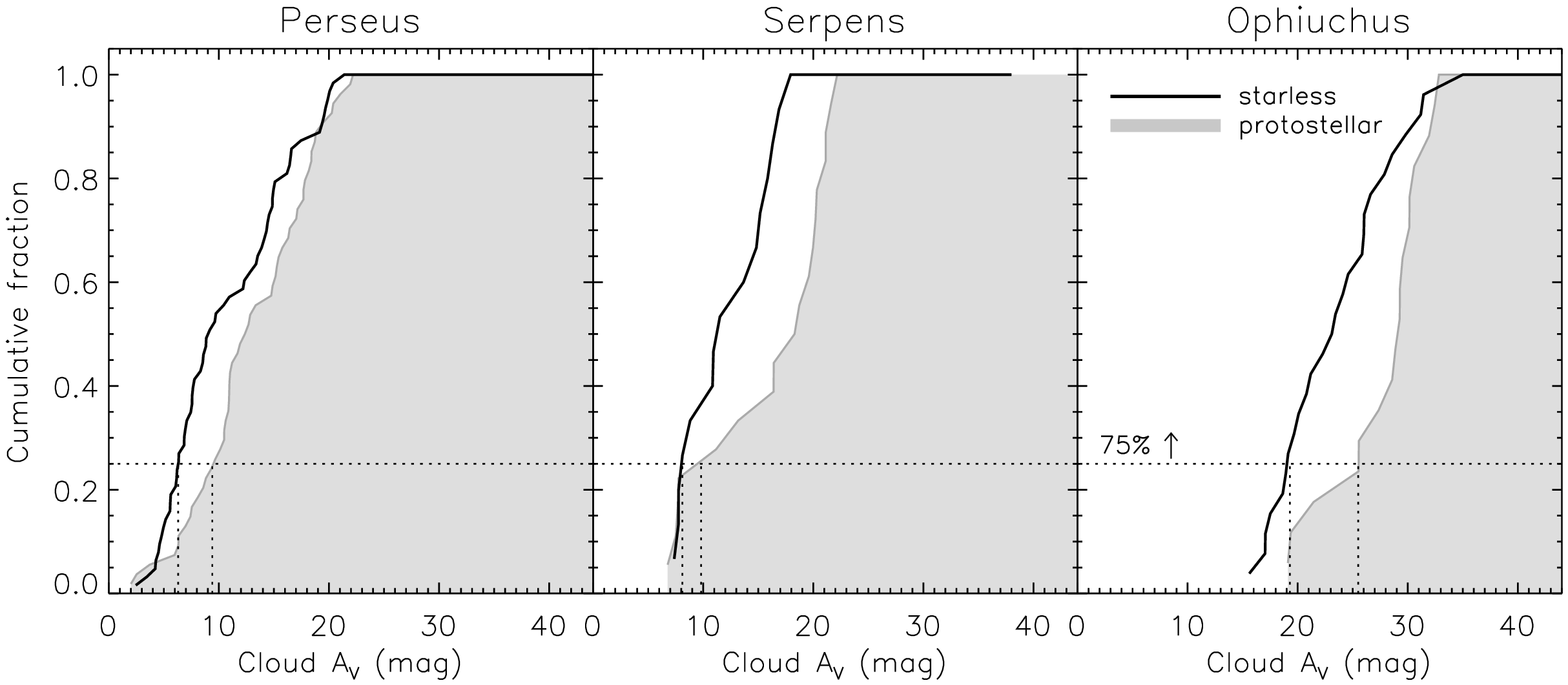}
\caption
{Cumulative fraction of starless and protostellar cores
  as a function of cloud \av, where the \av\ is derived from the
  reddening of background stars  in 2MASS and IRAC data.  In all three
  clouds, the majority of cores are found at high cloud column
  densities ($\av>7$~mag).  Dotted lines indicate the \av\ at which
  $>75\%$ of cores are found;   75\% of starless and protostellar
  cores in Perseus are located at $\av \gtrsim 6.5$ and 9.5~mag,
  respectively.  The equivalent values are $\av \gtrsim 6$ and
  10~mag in Serpens, and $\av \gtrsim 19.5$ and 25.5~mag in Ophiuchus.
  There appears to be a strict extinction threshold in Serpens and 
  Ophiuchus, with no cores found below $\av \sim 7$ and 15~mag, 
  respectively.}
\label{threshcomp}
\end{figure*}

We use the cumulative fraction of starless and protostellar cores as a
function of cloud \av, shown in Figure~\ref{threshcomp}, to quantify
the relationship between dense cores and the surrounding cloud
material.  Visual extinction (\av) is a  measure of the cloud column
density, and is  derived based on the reddening of background 2MASS
and IRAC sources, as described in Paper III and \citet{huard06}.  In
Paper III, we found that 75\% of 1.1~mm cores in Perseus, Serpens, and
Ophiuchus are found at visual extinctions of $\av \gtrsim 8$~mag, $\av
\gtrsim 15$~mag, and $\av \gtrsim 20-23$~mag, respectively.  Although
these values do not define a strict threshold, below these \av\ levels
the likelihood of finding an 1.1~mm core is very low.  Here we
investigate whether the relationship between dense cores and cloud
column density is different for starless and protostellar cores.  

Figure~\ref{threshcomp} demonstrates that both starless and
protostellar cores are found primarily at high cloud column densities
($\av>6$~mag):  75\% of starless and protostellar cores in Perseus are
located at $\av \gtrsim 6.5$ and 9.5~mag, respectively,  $\av \gtrsim
8$ and 10~mag in Serpens, and $\av \gtrsim 19.5$ and 25.5~mag in
Ophiuchus.   Note that these ``threshold'' values are significantly
higher than the  minimum cloud \av\, which is approximately 2~mag in
Perseus and Ophiuchus, and 6~mag in Serpens.  Starless cores tend to
be found at somewhat lower \av\ than protostellar cores; a two-sided
KS test yields probabilities of approximately 1\% that the starless
and protostellar distributions are drawn from the same parent
distribution in each cloud.  In Serpens and Ophiuchus there appear to
be strict extinction thresholds for 1.1~mm cores at 7 and 15~mag,
respectively.  A few cores in Perseus and Serpens lie outside the 
\av\ map area, and may be associated with low-\av\ material.

As discussed in Paper III, an extinction threshold has been
predicted by \citet{mckee89} for photoionization-regulated star
formation in magnetically supported clouds.  In this model, core
collapse and star formation will occur only in shielded  regions of a
molecular cloud where $\av \gtrsim 4-8$~mag.   The fact that 75\% of
both protostellar and starless cores are found  above $\av \sim 6$~mag in
each cloud is consistent with this model; while it is certainly not the only 
explanation, magnetic fields may play a role in inhibiting collapse 
of cores, at least in the low column density regions
of molecular clouds.

\begin{figure*}[!ht]
\includegraphics[angle=90,width=7.2in]{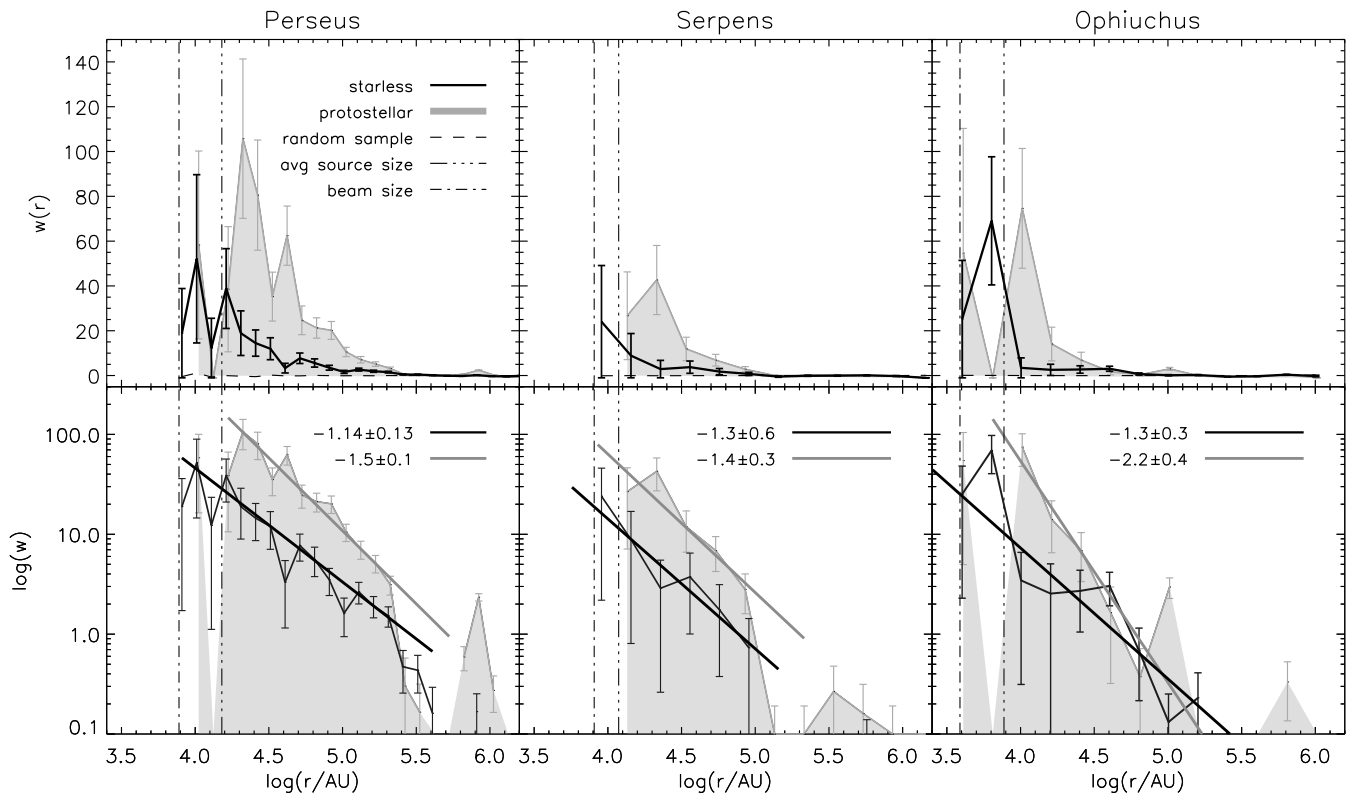}
\caption
{Two-point spatial correlation function $w(r)$, plotted as a function
  of source separation $r$,  for starless and protostellar cores in
  Perseus, Serpens, and Ophiuchus  (upper panels).  Lower panels plot
  log($w$), with the best fitting power law slope for each
  distribution. Resolution limits and average source sizes are indicated. 
  Power law slopes are fit for $r$ larger than the
  average source size in each cloud. The amplitude of the protostellar
  $w(r)$ is consistently higher than the starless $w(r)$, indicating
  that clustering of protostellar cores is stronger on all spatial
  scales.  Slopes for the starless and protostellar populations are
  similar in each cloud, however, so although the magnitude of
  clustering differs, the fundamental nature of the clustering does not.} 
\label{corrnfncomp}
\end{figure*}

\subsection{Clustering}\label{clustsec}

Finally, we look at the spatial clustering of starless and
protostellar cores. The spatial distributions of starless and
protostellar cores in each cloud are shown in Figure~\ref{spatial}.
Starless and protostellar cores appear to cluster together for the
most part, often congregating along filamentary structures in the
clouds.  Both populations occur in the known clusters (e.g. NGC 1333,
L 1688) and in a more distributed way across the clouds, with a few
exceptions, such as in Perseus where there is a group of primarily
starless cores (B1 ridge).

As a more quantitative measure of clustering, we use the  two-point
correlation function (Figure~\ref{corrnfncomp}):
\begin{equation}
w(r) = \frac{H_s(r)}{H_r(r)} - 1,
\end{equation}
where $H_s(r)$ is the number of core pairs with separation between
$log(r)$ and $log(r+dr)$,  and $H_r(r)$ is similar but for a random
distribution.  The correlation function $w(r)$ is a measure of the
excess clustering as compared to a random distribution of sources.
The upper panels of Figure~\ref{corrnfncomp} plot $w(r)$ as a function
of source separation $r$, with the linear beam size and average source
size indicated.  The lower panels plot log$(w)$, with a power law fit
($w(r) \propto r^p$) for $r$ larger than the average source size.  In
essence, the amplitude of $w(r)$ is a measure of the magnitude of
clustering, while the slope is a measure of how quickly clustering
falls off on increasing scales. 

The amplitude of $w(r)$ is higher for the protostellar samples  in all
three clouds, indicating that the degree of clustering is stronger on
all spatial scales for protostellar cores.   Visually, however,
starless and protostellar cores tend to cluster in a similar way
(Figure~\ref{spatial}), and this observation is supported by the
similarity in the slope of $w(r)$ for starless and protostellar cores.
In Serpens,  $p=-1.4\pm0.3$ and $-1.3\pm0.6$ for the starless and
protostellar populations, respectively.  The Ophiuchus curves are
quite noisy, but the best fit slopes ($-1.3\pm0.3$ and $-2.2\pm0.4$)
are consistent within $2\sigma$.  For Perseus the starless correlation
function is shallower by $4\sigma$ ($p=-1.14\pm0.13$ and
$-1.5\pm0.1$).  A shallower slope suggests that, while the amplitude
of clustering is weaker for starless cores, it does not fall off as
fast at larger spatial scales.  

A lower amplitude of clustering for starless cores as compared to
protostellar cores in Perseus and Serpens is confirmed by  the peak
number of cores per square parsec:  10~pc$^{-2}$ and 16~pc$^{-2}$,
respectively, for starless and protostellar cores in Perseus, and
4~pc$^{-2}$ and 8~pc$^{-2}$ in Serpens.  These numbers suggest that
clustering in the protostellar samples is a factor of 1.5--2 times
stronger than in the starless samples.  In Ophiuchus, however, the
values of the starless and protostellar populations are identical
(12~pc$^{-2}$).

There are at least three plausible reasons that clustering might be
stronger for protostellar cores, two environmental and one
evolutionary.  The difference may be an environmental effect if cores
that are located in regions of the cloud with higher gas density are
more likely to collapse to form protostars.  In that case, more
clustered sources would tend to be protostellar rather than starless.
Similarly, if outflows or other protostellar activity trigger the
collapse of nearby cores, we would again expect more protostellar
sources in clustered regions.  If evolution plays a more important
role, on the other hand, the spatial distribution of cores might
evolve after protostellar formation, for example as a result of
dynamical effects.

\section{Are the Starless Cores Really Prestellar?}\label{boundsec}

For any discussion of the mass distribution or
lifetime of prestellar cores, it is important to determine
whether or not our starless cores are likely to be truly prestellar
(i.e. will form one or more stars at some point in the future).
To this end, we investigate the dynamical state of starless cores 
by estimating the virial masses of cores in Perseus. 

While it is possible that some fraction of the observed starless cores
are transient or stable structures that will never form stars,   the
starless cores detected by our Bolocam surveys have high mean particle densities 
(typical $n_{1e4} \sim 1-3 \times 10^5$~cm$^{-3}$; \S\ref{densec}), 
making them  likely to be prestellar \citep{dif07,kc08}.     
Comparison of our data to molecular line observations is a more robust
method of determining if cores are gravitationally bound; line-widths
provide a direct estimate of the internal energy of cores, while the
1.1~mm dust masses provide an estimate of the potential energy.
An ammonia (NH$_3$) (J,K)=(1,1) and (2,2) survey of dense cores in
Perseus that includes all of the Bolocam-identified 1.1~mm cores has
recently been completed by \citet{ros07} at the GBT as part of the 
COMPLETE project.\footnote{``The COMPLETE Survey of Star Forming Regions''; 
http://cfa-www.harvard.edu/COMPLETE/ \citep{good04}.}
These observations provide a much-needed measure of the temperature
and internal motions of dense millimeter cores.

\begin{figure}
\hspace{-0.5in}
\includegraphics[width=4in]{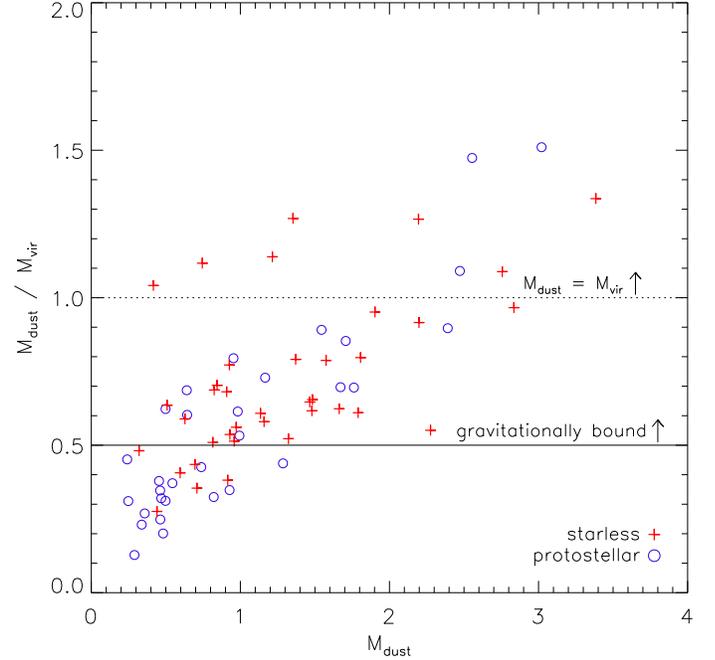}
\caption
{Ratio of the total mass derive from dust ($M_{dust}$) to the virial
  mass ($M_{vir}$),  calculated using the temperature and velocity
  dispersion derived from GBT NH$_3$ (1,1) and (2,2)  observations
  \citep{ros07}, for starless and protostellar cores in Perseus.  Only
  those with well-determined NH$_3$ line-widths are shown.   Solid and
  dashed lines indicate the minimum $M_{dust}/M_{vir}$ for which cores
  should be in virial equilibrium (dashed) or self-gravitating
  (solid).
\label{virmass}}
\end{figure}

Figure~\ref{virmass} plots the ratio of the dust mass
($M_{\mathrm{dust}}$, from Tables~\ref{sltab} and \ref{pstab})  to the
virial mass ($M_{\mathrm{vir}}$) as a function of $M_{\mathrm{dust}}$
for the 70 cores in Perseus that have well-determined line widths and
temperatures from  \citep{ros07} NH$_3$ survey.  Although all Bolocam
cores were detected in ammonia, not every target had a significant
NH$_3$ (2,2) detection or sufficient signal-to-noise to measure the
line width in the presence of the hyperfine structure.  
We present the data relevant to Figure~\ref{virmass} in Table \ref{vmtab}. 

\input{tab4} 

We estimate the virial mass of the cores as 
\begin{equation}
M_{\mathrm{vir}}= \frac{5 \sigma_{\mathrm{H2}}^2 R}{a G}, \label{vireq}
\end{equation}
where $a$ accounts for the central concentration of the core, $R$ is
the radius of the core, and $\sigma_{\mathrm{H2}}$ is the line width
of the molecular gas inferred from the ammonia line width.  
Note that $M_{\mathrm{dust}}/M_{\mathrm{vir}} =
1/\alpha$ for $a=1$, where $\alpha$ is the virial parameter introduced
by \citet{bm92}. 
For simplicity, we adopt a power-law density profile, for which
$a=(1-p/3)(1-2p/5)^{-1}$.  We report values in Table \ref{vmtab} for a
$p=1.5$ profile, giving $a=5/4$; for $p=2$ (SIS) the dynamical masses
would be 25\% smaller.   
The total line width of H$_2$ includes
thermal and turbulent components, and is calculated as
\begin{equation}
\sigma_{\mathrm{H2}}^2 =
\sigma^2_{\mathrm{NH3}}-\frac{kT_{K}}{17m_{\mathrm{H}}}+\frac{kT_{K}}{2m_{\mathrm{H}}}.
\end{equation}

The radius is estimated from the 1.1~mm data as $R=\theta_{dec}/2$, or
the HWHM.  Although the effective radius $R$ appropriate for the
virial theorem is not exactly equivalent to the gaussian-fit HWHM, we
do not expect the correction to be large.  If the FWHM is a good
representation of $\sqrt{8 \mathrm{ln}2}$ times the radial dispersion
one would measure in the x-y plane for a power law density profile,
then we expect
\begin{equation}
R = \theta_{dec}  \sqrt{(5-p)/(3-p)} \sqrt{3/4}
\sqrt{1/(8\mathrm{ln}2)}
\end{equation}
or $R \sim 0.56 ~\theta_{dec}$ for $p=1.5$, only a 10\% correction.

Dotted and solid lines in Figure~\ref{virmass} indicate regions in
the plot for which cores should be virialized and self-gravitating,
respectively.  Note here that $M_{\mathrm{dust}} = M_{\mathrm{vir}}$
implies $2K=-U$, where $K$ is the kinetic and $U$ the gravitational
energy, while the self-gravitating limit is defined by $K=-U$, or
$M_{\mathrm{dust}}/M_{\mathrm{vir}} = 0.5$.  There are two important
points to note here.  First, nearly all of the starless cores lie
close to or above the self-gravitating line, indicating that they are
likely to be gravitationally bound.  This statement must be modified
by the important caveat that the  power law profile assumed and the
method for measuring the dust mass introduce significant systematics
that could shift $M_{\mathrm{dust}}/M_{\mathrm{vir}}$ by more than a
factor of two.

Perhaps more convincingly, there is very little difference between
starless and protostellar cores in this plot; even though several of
the starless cores  lie below the self-gravitating line, this regions
is populated by protostellar  cores as well, which are by definition
capable of forming stars.  Furthermore, recent estimates of the dust 
opacity $\kappa_{1.1mm}$ (\citealt{shir07}; see \S~\ref{densec}) may increase 
$M_{\mathrm{dust}}$ by 1.3, making  the starless cores more bound.
Based on this discussion, we assume from here on
that all starless cores in our 1.1~mm samples
are true prestellar cores.

\section{The Prestellar Core Mass Distribution and the IMF}\label{slcmdsec}

One very important measure of the initial conditions of star formation
is the prestellar core mass distribution (CMD).  In particular,
comparing the  prestellar CMD to the stellar initial mass function
(IMF) provides insight into how the final masses of stars are
determined.    There are a number of processes that may (jointly)
dictate what the final mass of forming star will be.   We focus here
on two simple cases, assuming that only one drives the shape of the
IMF.  

If stellar masses are determined by the initial fragmentation
into cores, i.e., the final star or binary mass is always a fixed
percentage of the original core mass, then the shape of the emergent
stellar IMF should closely trace that of the prestellar CMD
\citep[e.g.,][]{myer98}.  This might be expected in crowded regions
where the mass reservoir is limited to a protostar's nascent core.
If, on the other hand, stellar masses are determined by competitive
accretion \citep{bon01}, or by the protostars themselves through
feedback mechanisms (e.g., outflows and winds; \citealt{shu87}), we
would not expect the emergent IMF to reflect the original core mass
distribution \citep{AF96}.

\subsection{The Prestellar CMD}\label{slcmdsec2}

\begin{figure*}
\includegraphics[width=7.3in]{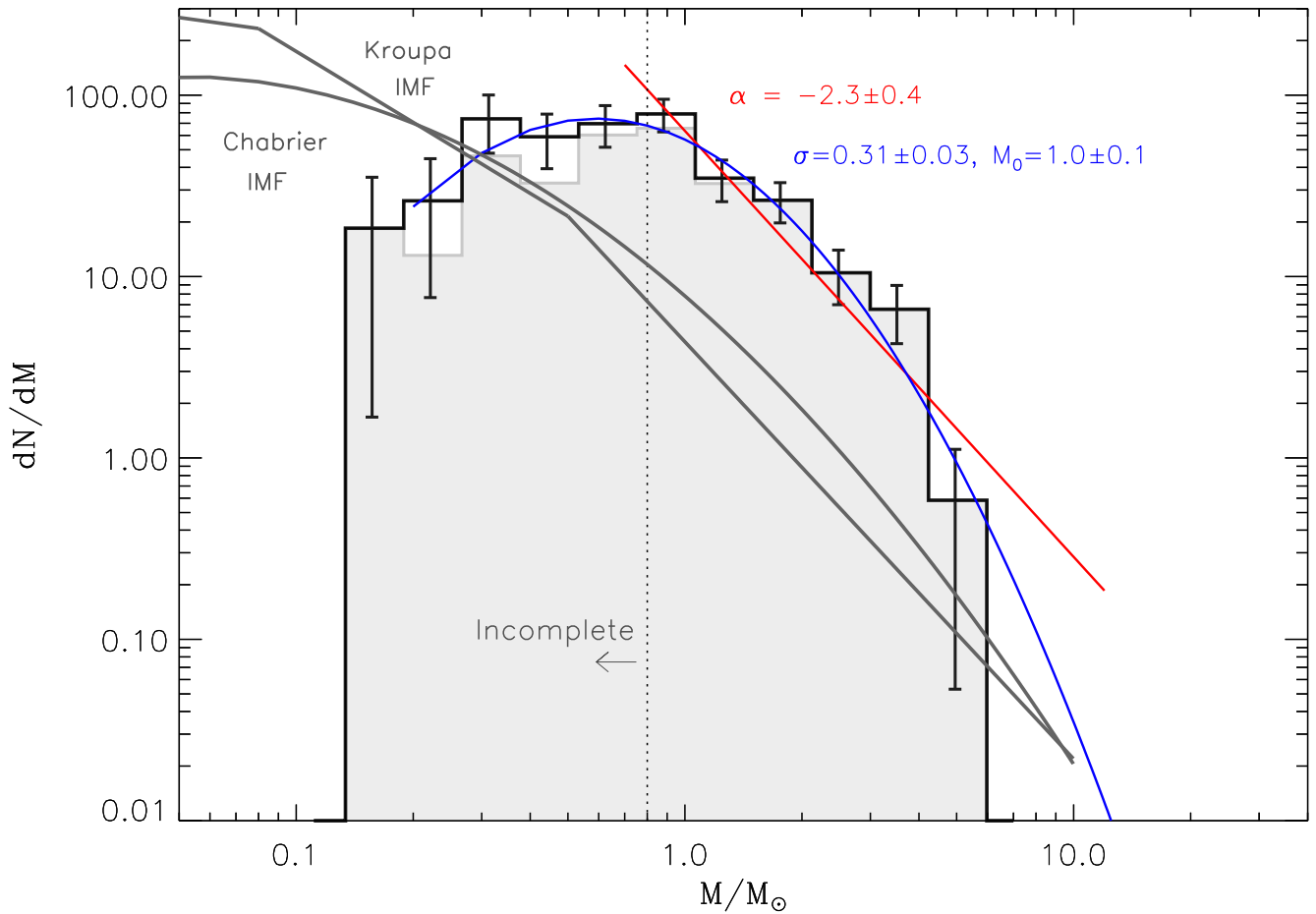}
\caption
{Combined prestellar core mass distribution (CMD), with power law and
  lognormal fits.  The prestellar sample is composed of all starless
  cores from Perseus, Serpens, and Ophiuchus, and the 50\% mass
  completeness limit (dotted line) is defined by the completeness
  limit for average-sized cores in Perseus.  Recent measurements of
  the stellar IMF for $M\gtrsim 0.5 \msun$ ($\alpha=-2.3$ to $-2.8$)
  are similar to the best-fit CMD power law slope
  ($\alpha=-2.3\pm0.4$).  IMF fits from \citet{chab05} (lognormal) and
  \citet{kroupa02} (three-component power law) are shown as thick gray
  lines for reference.  The shaded histogram indicates how the mass
  distribution changes if a small fraction of cores (15\%, based on
  the six ``unbound'' cores from Figure~\ref{virmass}) with $M<1
  \msun$ are excluded from the prestellar sample. 
\label{slmass}}
\end{figure*}

We combine the prestellar core samples  from all three clouds, 108
cores in total,  and assume $T_D=10$~K to calculate masses.   As noted
above in \S~\ref{boundsec}, our assumption that the majority of
starless cores in our sample are truly prestellar is supported by a
comparison of  the dust mass to the dynamical mass from NH$_3$
observations.   The resulting prestellar CMD is shown in
Figure~\ref{slmass}.  

The 50\% completeness limit shown ($M \sim 0.8~ \msun$;
dotted line) is estimated based on the empirical 50\% completeness curve in
Figure~\ref{mvscomp}.  Our completeness depends on the size
of a given source; as the average core size in
Perseus is $68\arcsec$, we take the 50\% completeness
limit for a 70\arcsec\ FWHM source ($M\sim 0.8 ~\msun$) to be the average
50\% completeness limit of the sample.  The equivalent 90\% completeness
limit is a factor of 1.2 higher, $M\sim 0.93 ~\msun$.  50\% completeness
limits for average-sized sources are lower in Serpens ($0.6~ \msun$)
and Ophiuchus ($0.5~ \msun$), but as more than half of the total
population of prestellar cores are in Perseus, we take $0.8~ \msun$
for the entire prestellar sample.  

We fit a power law ($dN/dM \propto M^{\alpha}$) to the CMD for $M>0.8
~ \msun$, finding a slope of $\alpha=-2.3$, with a reduced chi-squared
of $\tilde{\chi}^2=1.9$.   The best fit slope depends somewhat on the
histogram binning, ranging  from 2.0 to 2.6 for bin widths of 0.1 to
0.3 $M_{\sun}$, so we assign a  total uncertainty of $0.4$ to our best
fit slope, when also taking into account  formal fitting errors.  We
fit a lognormal distribution  to $M>0.3 ~ \msun$, finding a best-fit
width $\sigma=0.30\pm0.03$ and characteristic mass $M_0 = 1.0 \pm 0.1
~ \msun$.  Although the lognormal function is quite a good fit
($\tilde{\chi}^2=0.5$), the reliability of the turnover in the
prestellar CMD is highly questionable given that the completeness
limit in Perseus coincides closely with the turnover mass.  The
prestellar CMD can also be fit by a broken power law with $\alpha=-4.3
\pm1.1$ for $M>2.5~ \msun$ and $\alpha=-1.7 \pm0.3$ for $M<2.5~
\msun$, although the uncertainties are large. 

Given that source detection is based on peak intensity, 
we may be incomplete to sources with very large sizes and 
low surface density even in the higher mass bins ($M>0.8\msun$).   
Completeness varies with
size similarly to $M\propto R^2$, thus the fraction of (possibly) missed
sources should decrease with increasing mass.\footnote{Unless $M\propto
  R^2$ intrinsically for starless cores, in which case a constant
  incompleteness fraction would apply over all mass bins.} 
The effect of missing such low surface brightness cores, if they exist and
could be considered prestellar, would be to flatten the CMD slightly
(i.e. the true slope would be steeper than the observed slope). 
Instrumental selection effects are discussed further in Paper~I.

To be completely consistent, we should exclude the ``unbound'' cores
from Figure~\ref{virmass} (those with
$M_{\mathrm{dust}}/M_{\mathrm{vir}} < 0.5$) from our prestellar CMD.
This represents 6 out of the 40 cores that have measured virial masses in
Perseus, all 6 of which have $M_{\mathrm{dust}}<1\msun$.  We do not have
virial masses for cores in Serpens or Ophiuchus, but we can randomly
remove a similar fraction of sources with  $M<1\msun$ from each cloud
sample (2 sources from Serpens, 4 from Ophiuchus, and an additional 4
from Perseus).  The shaded histogram in Figure~\ref{slmass}
indicates how the mass distribution is altered when these 16
``unbound'' cores are excluded from the sample.  Our derived CMD slope
is not affected, as  nearly all of the starless cores below the
``gravitationally bound''  line in Figure~\ref{virmass} have masses
below our completeness limit, and even at low masses the CMD is not
significantly changed.

\begin{figure}
\hspace{-0.35cm}
\includegraphics[width=3.7in]{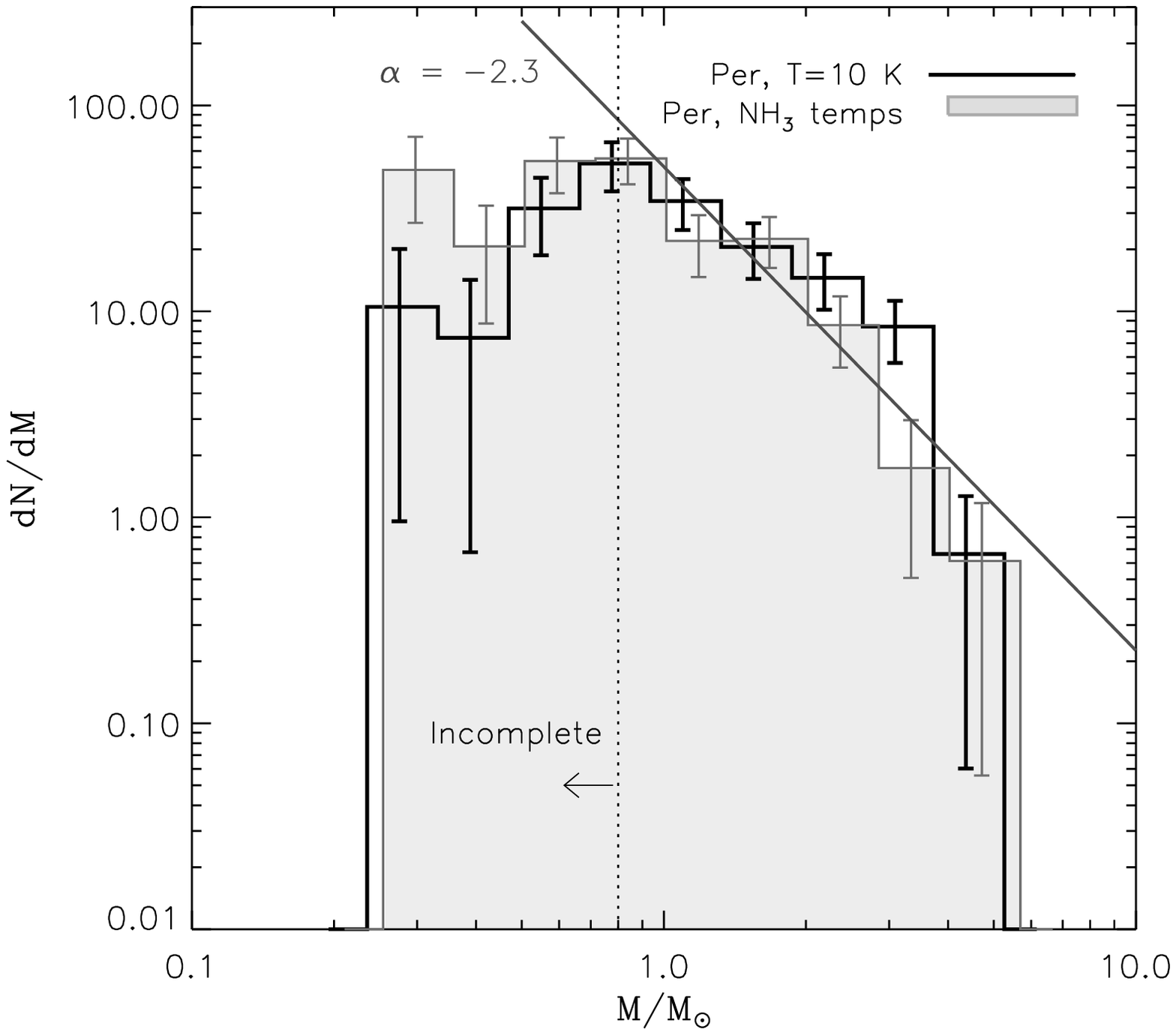}
\caption
{Effect on the prestellar CMD in Perseus of using NH$_3$-derived core
  kinetic temperatures (shaded histogram) rather than assuming a
  constant $T_D=10 K$ for all cores (black histogram).  Only starless
  cores in Perseus are included here.  The best-fitting power-law
  slope for $M>0.8 \msun$ is similar for the two histograms:
  $\alpha=-2.35\pm0.33$ and $\alpha=-2.33\pm0.28$ for the $T_D=10$~K 
  and NH$_3$ temperature curves, respectively.
  Although we do not have kinetic
  temperatures for cores in Serpens or Ophiuchus, we do not expect
  temperature effects to dramatically alter the derived prestellar CMD
  slope.
\label{ttemp}}
\end{figure}

There may also be some concern over the use  of a single dust
temperature $T_D=10$~K for all cores.  To test the validity of this
assumption, we use the kinetic temperatures ($T_K$) derived from the
GBT NH$_3$ survey of Perseus (\citealt{ros07}, Schnee et al. 2008, in
prep) to compute core masses, assuming that the dust and gas are well
coupled (i.e., $T_D=T_K$).   Figure~\ref{ttemp} shows the CMD of
prestellar cores in Perseus, both for a single $T_D=10$~K, and for
masses calculated using the NH$_3$ kinetic temperatures for each core.
There is some change to the shape of the CMD at intermediate masses,
but the best-fitting slope for $M>0.8\msun$ ($\alpha=-2.3$) is
unchanged.  In fact, the deviation from a power law is smaller when
using the kinetic temperatures than when using $T_D=10$ K ($\tilde
\chi^2=1.3$ and 2.5, respectively).  

The median $T_K$ of prestellar cores in Perseus is 10.8~K, quite close
to our adopted $T_D=10$~K, and the small overall shift in masses 
(a factor of 1.1) corresponding to an 0.8~K temperature difference 
would not affect the derived CMD slope.  The dispersion in kinetic
temperatures in Perseus is $\pm 2.4$~K, or $\pm 0.4 \msun$ for a $1
\msun$ core, and the tail of the distribution extends to $T_K > 15$~K
(Schnee et al. 2008, in prep).  We do  not have temperature
information for cores in Serpens or Ophiuchus; if the median
temperature were to vary from cloud to cloud by more than a few K, the
shape of the combined CMD could be significantly altered.  Perhaps a
more serious issue is that prestellar cores are not in reality
isothermal, but taking into account variations of the temperature with
radius requires radiative transfer modeling of each source.

\subsection{Comparison to the IMF}

The shape of the local IMF is still uncertain \citep{scalo05}, but
recent work has found evidence for a slope of $\alpha=-2.3$ to $-2.8$
for  stellar masses $M \gtrsim 1 ~ \msun$, similar to the slope we
measure for the combined prestellar CMD ($\alpha=-2.3\pm0.4$).  For example,
\citet{reid02} find $\alpha = -2.5$ above 0.6 ~ \msun, and $\alpha =
-2.8$ above 1~\msun.
\citet{schroder03} suggest $\alpha = -2.7$ for
$1.1 < M < 1.6$  \msun\ and $\alpha = -3.1$ for $1.6 < M < 4$ \msun.
For reference, the Salpeter IMF has a slope of $\alpha=-2.35$
\citep{sal55}, and the \citet{scalo86} slope  for sources with mass
$M\gtrsim 1~ \msun$ is $\alpha \sim -2.7$.  At lower masses, the IMF
flattens, and may be characterized by a lognormal function.
\citet{kroupa02} suggests a three-component power law for the 
average single-star IMF: $\alpha=-2.3$ for $0.5<M<1 ~ \msun$,
$\alpha=-1.3$  for $0.08<M<0.5 ~ \msun$, and  $\alpha=-0.3$ for
$0.01<M<0.08 ~ \msun$.
\citet{chab05} finds that a lognormal distribution with $\sigma=0.55$ 
and $M_0= 0.25~M_{\sun}$ is a good fit for $M<1~ \msun$.  

The \citet{kroupa02} three-component power law and the \citet{chab05}
lognormal IMFs are shown as thick gray lines in Figure~\ref{slmass}.
The width of the \citet{chab05} lognormal ($\sigma=0.55$) is somewhat
larger than the width of the prestellar CMD best-fit lognormal
($\sigma=0.3$), as expected if we are incomplete at lower masses, and
the IMF characteristic mass ($M_0=0.25~\msun$) is a factor of four
lower than that of the CMD ($M_0=1.0~\msun$).  A lower characteristic
mass for the IMF is expected if some fraction of the core mass is lost
in the star-formation process.  Power law fits for $M>1~ \msun$ appear
to be quite similar for the CMD ($\alpha=-2.3\pm0.4$) and IMF
($\alpha=-2.3$ to $ -2.8$), however.  

Although we cannot rule out the importance of feedback and other local
processes in determining the shape of the IMF, the fact that the
prestellar CMD and the local IMF have similar shapes 
supports a growing body of evidence that the final
masses of stars are determined during the core formation process.  If
this is the case, it is tempting to relate both the prestellar CMD and
the IMF to the CMD created from turbulent simulations.  Numerical
models of turbulent clouds have had some success in reproducing the
general shape of the IMF \citep[e.g.,][]{pn02,li04}, but the link is
not firmly established.  For example, we found in Paper III that the
predicted dependence of CMD shape on the turbulent mach number
\citep{paredes06,pn02} does not agree with observations.

Evidence for a direct link between the CMD and the IMF has been found
previously based on dust emission surveys of small regions
\citep{ts98,man98}, as well as molecular line observations of dense
cores (\citealt{oni02}).  Recently, \citet{alves07} found evidence for
flattening at low masses in the CMD of the Pipe Nebula, as traced by
dust extinction toward background stars.  Those authors interpret the
similarity between the Pipe Nebula CMD and the Trapezium cluster IMF
\citep{muench02} as evidence that the stellar IMF is a direct product
of the CMD, with a uniform core-to-star efficiency of $30\%\pm10\%$.
Although the measured masses of \citet{alves07} are somewhat less
uncertain than ours because they do not need to assume a dust opacity
or temperature, the mean particle densities of cores traced by dust extinction
($n\sim 5\times10^3 - 2\times10^4$~cm$^{-3}$) are considerable lower
than the mean densities of cores traced by our Bolocam 1.1~mm surveys
($n \sim 2\times 10^4 - 10^6$~cm$^{-3}$), and they may never form
stars.  In fact, recent C$^{18}$O and NH$_3$ observations suggest that
the dust extinction sources  in the Pipe Nebula are pressure confined,
gravitationally unbound starless cores \citep{lada07,muench07}.  In
Orion, \citet{nw07} find a turnover in the CMD of starless SCUBA
$850\micron$ cores at $\sim 1.3 \msun$.  Those authors relate this
turnover to a down-turn in the \citet{kroupa02} IMF at $\sim 0.1
\msun$, and infer a much lower core-to-star efficiency of $\sim 6\%$.

If the prestellar CMD does have a one-to-one relationship with the
stellar IMF, then the ratio of turnover masses is a measure of the core collapse efficiency, 
or the fraction of original core mass that ends up
in the final star:  $f_{\mathrm{eff}}=M_{\mathrm{TO}}^{\mathrm{IMF}}/M_{\mathrm{TO}}^{\mathrm{CMD}}$.
Here $M_{\mathrm{TO}}$ is the mass where the $dN/dM$ distribution,
which rises with decreasing mass, flattens out and begins to fall with
further decreasing mass. 
Equivalently,
$M_{\mathrm{TO}}^{\mathrm{CMD}}/M_{\mathrm{TO}}^{\mathrm{IMF}} =
1-f_{\mathrm{eff}}$ is the fraction of core mass lost in the star
formation process.   

Our limited completeness dictates that we can only measure a lower
limit to $f_{\mathrm{eff}}$.  If there is a true turnover in the
prestellar CMD, it must occur below our completeness limit, at
$M_{\mathrm{TO}}^{\mathrm{CMD}} \lesssim 1.0~ \msun$.   The system
IMF, i.e., treating binaries and multiple systems as single rather
than multiple objects, peaks at $M_{\mathrm{TO}}^{\mathrm{IMF}} \sim
0.2-0.3~ \msun$ \citep[e.g.,][]{chab05,luhm03}.  This system IMF is
appropriate for comparison to our CMDs, as we would not resolve such
multiple systems even if they form from distinct cores.  For an IMF
turnover mass of $0.25~ \msun$, $M_{\mathrm{TO}}^{\mathrm{CMD}}
\lesssim 1.0~ \msun$ implies that at least 25\% of the initial core
mass is accreted onto the final star or stellar system.
Characteristic masses associated with lognormal fits to both the IMF
and CMD imply a similar ratio of $f_{\mathrm{eff}} \gtrsim
M_0^{\mathrm{IMF}}/M_0^{\mathrm{CMD}} \sim 0.25/1.0 \sim 0.25$.   

Our conclusion that the core-to-star efficiency is at least 25\% is
consistent with the value found by \citet{alves07} for the Pipe Nebula
(30\%),  and with predicted efficiencies of $25\%-75\%$ from recent
analytic models of bipolar protostellar outflows \citep{matz00}.

\subsection{The Effect of ``Multiple'' Sources}\label{multsec}

As noted in \S~\ref{embsl}, $17\%-55$\% of our protostellar cores are
associated with more than one (two to three) embedded protostars.  We
can expect that a similar fraction of prestellar cores will form a
resolved binary or multiple system, making a direct mapping between
the CMD and IMF difficult to justify unless the resolutions are
matched.  Given the relatively low resolution of the Bolocam data
($30\arcsec$), some cores must result in wide-separation multiple
stellar systems that would not be considered single objects in the
system IMF.  Furthermore, the system IMF likely evolves over time 
due to dynamical effects such as the decay of high-order multiples, 
ejections, and close interactions \citep[see][for more 
discussion of the CMD to IMF mapping]{good07}.

The effect of these multiple sources will likely be to steepen the
CMD, as higher mass cores are divided into multiple lower mass
sources.  This is consistent with the fact that the prestellar CMD
slope is at the low end of the IMF slope range.  It is important to
keep in mind that we would expect only $\sim 25-30 \%$ of cores in 
the combined prestellar CMD 
to be ``multiple'' sources, so they should not dominate the CMD slope.
While we know that most stars occur in binaries 
\citep[][and references therein]{duch07}, and thus that
there must be some unresolved binaries in the \textit{Spitzer} data,
these compact systems will most likely be unresolved in the system IMF
as well, so they do not affect our comparison.

\begin{figure*}[!ht]
\centering
\includegraphics[width=6.in]{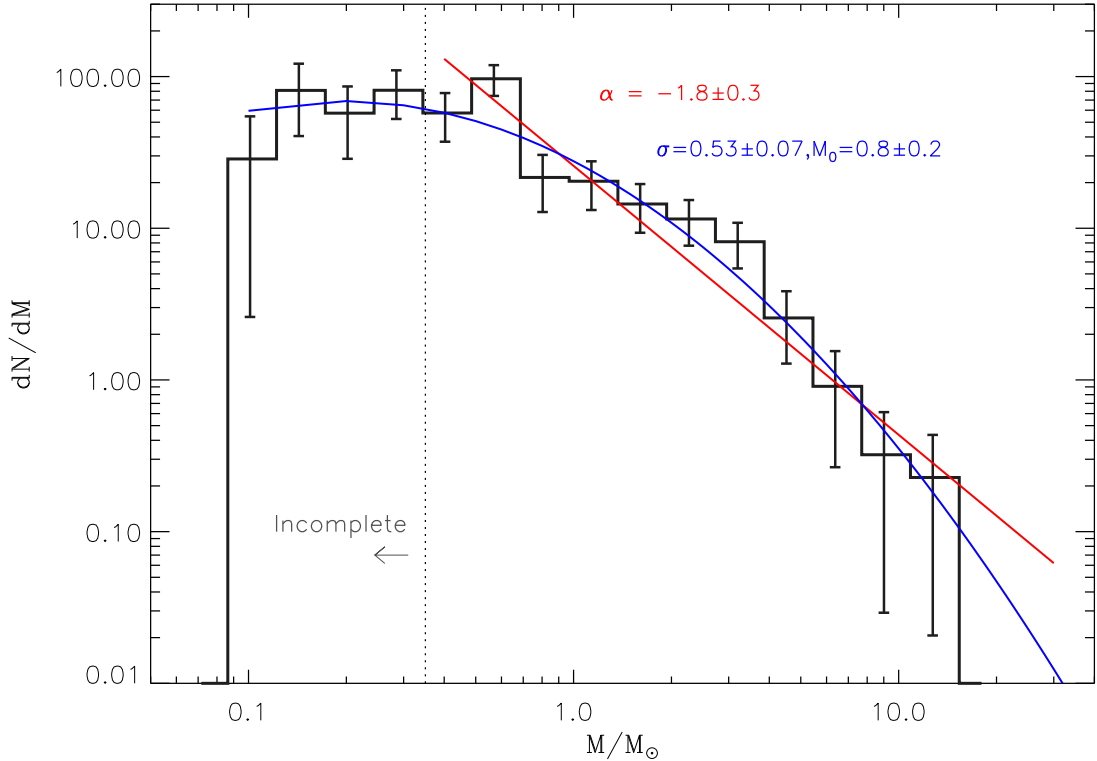}
\caption
{Combined protostellar mass distribution, with power law and
  lognormal fits.  The protostellar CMD includes all
  protostellar cores from Perseus, Serpens, and Ophiuchus, and the
  50\% mass completeness limit is defined by the completeness to
  average sized protostellar cores in Perseus.  The best-fit power law
  ($\alpha=-1.8$) is shallower, and the distribution as a whole is
  wider, than the prestellar CMD (Figure~\ref{slmass}).  This
  difference is expected if the protostellar CMD evolved from the
  prestellar CMD.  The fact that the protostellar distribution
  extends to higher masses than the prestellar CMD is more difficult
  to explain, however, and may suggest that higher mass prestellar cores
  are relatively short lived.  
\label{psmass}}
\end{figure*}

\subsection{The Protostellar Core Mass Distribution}\label{pscmdsec}

For comparison, we show in Figure~\ref{psmass} the mass distribution
of the combined \textit{protostellar} core sample from all three
clouds,  assuming $T_D = 15$~K.  The protostellar CMD is considerably
wider and flatter than the prestellar CMD, and extends to higher
masses.    The best-fitting power law slope ($\alpha=-1.8$;
$\tilde{\chi}^2=1.7$) is shallower than for the prestellar sample, and
the best-fitting lognormal distribution ($\sigma=0.51\pm0.07$, $M_0 =
0.8\pm0.2$; $\tilde{\chi}^2=1.3$) is wider by nearly a factor of two.
The protostellar CMD must be interpreted in the context of a constant
assumed $T_D = 15$~K; an intrinsic spread of temperatures would tend to
flatten the mass distribution even more, but 15~K has been determined
to be the best single $T_D$ based on many detailed radiative transfer
models \citep{shir02,young03}.

A protostellar CMD that extends to lower masses than the prestellar
CMD is not unexpected, given that some fraction of the core mass has
already been accreted on to the central source for protostellar cores.
An extension to higher masses  in the protostellar CMD as compared to
the prestellar CMD, however, can only be explained if (a) we are
underestimating the dust temperature of protostellar sources or
underestimating the total  mass of prestellar cores, (b) the current
populations of prestellar cores in these clouds have lower mass than
the generation of cores that have already formed protostars, or (c)
the highest mass prestellar cores collapse to form protostars very
quickly.

\citet{hatch08} examine the relationship between starless and
protostellar mass distributions in Perseus based on SCUBA data,
finding that a simple mass-dependent evolutionary model in which
higher mass cores form protostars on a shorter timescale than lower
mass cores can qualitatively explain the observed steeper prestellar
CMD slope, and extension of the protostellar CMD to higher masses. 
Another step toward resolving the reasons for these differences may be
careful radiative transfer models of each prestellar and protostellar
source to replace the assumption of a constant $T_D$.

\section{Lifetime of the Prestellar Phase}\label{lifesec}

We can use the relative  number of prestellar
cores and embedded protostars to estimate the lifetime of the
prestellar  core phase, which is essential for understanding the
physical processes that govern the formation, support, and collapse of
star-forming cores.  Several assumptions are required for this
calculation.  We assume that all of the starless cores in our sample
are in fact prestellar, and will eventually collapse to form stars or
brown dwarfs (see \S\ref{boundsec}).  As we ultimately calibrate our
lifetimes based on the Class~II phase ($2 \times
10^6$~yr; \citealt{ken90,cieza06,spezzi08}), star formation must
have been  steady in time for at least the last 2~Myr.  In addition,
we must assume that  there is no significant evolutionary dependence
on source mass.   Although this kind of analysis relies on a number of
assumptions,  it has the advantage of being quite simple, and it does
not require on an a priori accretion rate or star-formation model.  

Table~\ref{lifetab} lists the  ratio of the number of starless cores
($N_{\mathrm{SL}}$) to embedded protostars ($N_{\mathrm{emb}} = N_{\mathrm{Class 0}}+N_{\mathrm{Class I}}$)
for all three clouds, and the starless core lifetime ($t_{\mathrm{SL}}$)
derived from that ratio:  $t_{\mathrm{SL}} = t_{\mathrm{emb}} (N_{\mathrm{SL}}/N_{\mathrm{emb}})$.  
The number of embedded protostars in each
cloud is derived in a companion paper (Enoch et al. 2008, in prep), in
which the SED derived from c2d \textit{Spitzer} photometry and 
1.1~mm fluxes is used to classify protostars based on their bolometric
temperature, $T_{bol}$: Class~0 ($T_{bol}<70$~K) and Class~I ($70
<T_{bol}<650$~K). In addition to the $T_{bol}$ classification, all
embedded protostars  are required to be detected at 1.1~mm, so that we
are complete to embedded sources with $M_{env} \gtrsim 0.1 \msun$
(Enoch et al. 2008, in prep).

\begin{deluxetable*}{llcccc}
\tablecolumns{6}
\tablewidth{0pc}
\tablecaption{\label{lifetab}Lifetime of the prestellar core phase}
\tablehead{
\colhead{Cloud} &  \colhead{$\mathrm{N_{\mathrm{SL}}/N_{\mathrm{emb}}}$} &  \colhead{$t_{\mathrm{\mathrm{SL}}}$}   &  \colhead{$\mean{n_{1e4}}_{\mathrm{\mathrm{SL}}}$}   &  \colhead{$t_{\mathrm{\mathrm{ff}}}$}  & \colhead{$t_{\mathrm{\mathrm{SL}}} / t_{\mathrm{\mathrm{ff}}} $}  \\
\colhead{} & \colhead{} & \colhead{(yr)} &  \colhead{(cm$^{-3}$)} & \colhead{(yr)} & \colhead{} 
}
\startdata
Perseus   &   $67/66=1.0$   &  $5\times10^5$  &  $1.4\times10^5$   &    $1.4\times10^5$   &  3.9  \\ 
Serpens   &   $15/35=0.4$   &  $2\times10^5$  &  $1.2\times10^5$   &    $1.5\times10^5$   &  1.5  \\ 
Ophiuchus &   $26/28=0.9$   &  $5\times10^5$  &  $1.6\times10^5$   &    $1.3\times10^5$   &  3.9  \\
 &     & &   &   &   \\
Combined sample & $108/129=0.8$ &  $4.5\times10^5$  &  $1.7\times10^5$   &    $1.4\times10^5$   &  3.2  \\
\hspace{0.25cm} ``low density''  & $78/129=0.6$ &  $3.3\times10^5$  &  $1.3\times10^5$   &   $1.5\times10^5$   &  2.2 \\  
\hspace{0.25cm}  ``high density''  & $30/129=0.2$ &  $1.3\times10^5$  &  $2.8\times10^5$   &   $1.0\times10^5$   &  1.3   \\
\enddata
\tablecomments{$\mathrm{N_{\mathrm{emb}}}$ refers to the total number of embedded protostars, i.e., those in the Class~0 and Class~I phases: $\mathrm{N_{\mathrm{emb}} = N_{\mathrm{Class 0}} + N_{\mathrm{Class I}}}$. The number of embedded protostars is derived in a companion paper (Enoch et al. 2008, in prep).  The prestellar core lifetime $t_{\mathrm{SL}}$ is derived from the number of starless cores:  $t_{\mathrm{SL}} = t_{\mathrm{emb}} (N_{\mathrm{SL}}/N_{\mathrm{emb}})$, for an embedded phase lifetime $t_{\mathrm{emb}} = 5.4 \times 10^5$~yr (Evans et al. 2008, in prep).  Core mean densities $n_{1e4}$ are calculated in a fixed linear aperture of diameter $10^4$~AU, and the free-fall timescale $t_{\mathrm{ff}}$ of starless cores is derived from the typical mean density $\mean{n_{1e4}}_{\mathrm{SL}}$ of the starless samples using Eq.~(\ref{tffeq}).  The ``Combined sample'' includes sources from all three clouds; this combined sample is divided into ``low density'' and ``high density'' bins representing cores with $n_{1e4} < 2\times 10^5$ cm$^{-3}$ and $n_{1e4} > 2\times 10^5$ cm$^{-3}$, respectively.}
\end{deluxetable*}

The ratio of the number of starless cores to the number of embedded
protostars is $\mathrm{N_{\mathrm{SL}}/N_{\mathrm{emb}}}=1.0$ in Perseus, 0.4 in
Serpens, and 0.8 in Ophiuchus.   For an embedded protostar phase that
lasts $t_{\mathrm{emb}} = 5.4 \times 10^5$~yr (Evans et al. 2008, in prep), the
observed ratios imply prestellar core  lifetimes of $5 \times 10^5$~yr
in Perseus, $2 \times 10^5$~yr in Serpens, and $5 \times 10^5$~yr in
Ophiuchus.  
Note that we also found approximately equal numbers of
protostellar and starless cores in all three clouds
(Table~\ref{coretab}), further confirmation that the lifetime of
starless cores is similar to that of the embedded protostellar
phase.\footnote{The $\mathrm{N_{\mathrm{SL}}/N_{\mathrm{emb}}}$ ratios in Table~\ref{lifetab} differ
  from $\mathrm{N_{\mathrm{SL}}/N_{\mathrm{PS}}}$ in Table~\ref{coretab}
  due to the presence of multiple embedded
  protostars in some protostellar cores (\S\ref{embsl}) and to the
  fact that some embedded protostars are ``band-filled'' at 1.1~mm.
  Band-filled sources are not associated with a distinct core, but
  appear to be associated with millimeter emission that is either
  extended or below the $5\sigma$ detection limit.}  Therefore, the
dense starless cores we are sensitive to last for $2-5 \times10^5$~yr
in all three clouds.  The differences in $\mathrm{N_{\mathrm{SL}}/N_{\mathrm{emb}}}$
from cloud to cloud may be environmental.  The ratio is lowest in
Serpens, which has the highest mass cores on average.  As noted in
\S\ref{pscmdsec}, higher mass cores tend to be protostellar, possibly
because they collapse on a shorter timescale than lower mass cores.

Taking all three clouds together yields a prestellar core lifetime of
$4.5 \times 10^5$~yr.  The uncertainty in the measured number of
prestellar cores is approximately $\pm 10$, based on the range in the
number of starless cores for different identification criteria (see
\S~\ref{assocsec}).  Given a similar uncertainty in the number of
embedded protostars (see Enoch et al. 2008, in prep.), this  corresponds to
an uncertainty in the lifetime of prestellar cores of approximately
$0.8 \times 10^5$~yr.

Published measurements of the prestellar core lifetime vary by  two
orders of magnitude, from a few times $10^5$~yr to $10^7$~yr
\citep{wt07}.  For example, \citet{lm99} calculate a core lifetime of
$6\times10^5$~yr for optically selected cores with mean densities of
$6-8\times10^3$~cm$^{-3}$, while \citet{jw00} find a lifetime of
$10^7$~yr for low density cores detected from column density maps
based on IRAS far-infrared observations.     Our results are similar
to recent  findings in Perseus by \citet{jorg07} and \citet{hatch07},
both of whom find approximately equal lifetimes for the starless and
embedded protostellar phases  by comparing SCUBA $850~\micron$ maps
with \textit{Spitzer} c2d data.  The approximate equality between
$N_{\mathrm{SL}}$ and $N_{\mathrm{PS}}$ has been  found by a number of studies, and was
noted early on by \citet{beich86}  for the NH$_3$ cores of
\citet{mb83}.   Our results are also consistent within a factor of two
with the lifetime derived by \citet{vrc02} for a sample of Lynds dark
clouds observed with SCUBA.

The average mean density (calculated in a fixed linear aperture of
$10^4$~AU, see \S~\ref{densec})  of the starless core samples,
$\mean{n_{1e4}}_{\mathrm{SL}}$, and corresponding free-fall timescale,
$t_{\mathrm{ff}}$, are also given in Table~\ref{lifetab}.   The
free-fall time is the timescale on which starless cores will collapse
in the absence of internal support, and is calculated from the mean
particle density \citep{spitz78}:
\begin{equation} 
t_{\mathrm{ff}} = \sqrt{\frac{3\pi}{32 G \rho}} = \sqrt{\frac{3\pi}{32 G
    \mean{n} \mu_p m_H}},\label{tffeq}
\end{equation} 
where $m_H$ is the mass of hydrogen and $\mu_p = 2.33$ is the mean
molecular weight per particle.  Mean densities are similar in all
three clouds, $\mean{n_{1e4}}_{\mathrm{SL}} = 1-2\times10^5$~cm$^{-3}$, with
corresponding free-fall times of $t_{\mathrm{ff}} = 1.3-1.5 \times 10^5$~yr.
The final column in Table~\ref{lifetab} gives the ratio of the
measured starless core lifetime to the average free-fall timescale in
each cloud: $t_{\mathrm{SL}}/t_{\mathrm{ff}} = 1.5-3.9$.  Thus prestellar cores last
for only a few free-fall times in all three clouds.   

Such a short prestellar core lifetime argues for a dynamic, rather
than quasi-static, core evolutionary scenario.  One classical
quasi-static model is that of magnetically dominated star formation,
in which the evolution of highly sub-critical cores is moderated by
ambipolar diffusion \citep{shu87}.  In this paradigm, prestellar cores
should have lifetimes similar to the ambipolar diffusion timescale, or
$t_{\mathrm{AD}} \sim 7\times 10^6$~yr for  typical ionization levels in
low-mass star forming regions \citep[e.g.,][]{nak98,evans99}, more
than an order of magnitude longer than our results.   

It is important to emphasize here that our Bolocam surveys are
sensitive to to  cores with relatively high mean density ($n\gtrsim
2-3 \times 10^4$~cm$^{-3}$; Paper III).\footnote{Note that this
  limiting density takes into account the 50\% completeness limit to
  starless cores as discussed in \S\ref{mvssec} and \S\ref{slcmdsec2};
  it is the typical density found when one calculates a mean density
  along the 50\% completeness curve as a function of size in
  Figures~\ref{mvscomp} and ~\ref{mvscomp2}.}  Thus  we may be
sampling only the densest end stage in a longer core evolutionary
picture, in which case a magnetic field dominated scenario could still
be applicable at early times (e.g., \citealt{tm04}).  \citet{cb01}
also note that $t_{AD}$ can be as short as a few  free-fall times for
marginally sub-critical cores.

\subsection{Density Dependence}

Interestingly, if we divide our sample into two density bins, $n_{1e4}
< 2\times 10^5$ cm$^{-3}$ and $n_{1e4} > 2\times 10^5$ cm$^{-3}$, we
find a longer lifetime for lower density cores: $3.3\times 10^5$ yr
versus $1.3 \times 10^5$ yr for the higher  density sample (see
Table~\ref{lifetab}).  This result suggests that the prestellar core
lifetime becomes shorter as cores become more centrally condensed, at
a faster rate than $n^{-0.5}$ (i.e., that expected based on the
dependence of $t_{\mathrm{ff}}$ on $n$, see Eq.\ref{tffeq}).   The
observed trend, $t_{\mathrm{SL}} \propto n^{-1.2}$, is closer to the
$n^{-0.85}$  dependence suggested by \citet{jw00}.

\subsection{Additional Uncertainties}

The prestellar lifetime depends on the ratio with protostellar cores,
but our sensitivities to prestellar and protostellar sources are not
equal due to the different assumed dust temperatures: the
protostellar core mass completeness limit is lower by approximately a
factor of two.  Of the 55 protostellar cores in Perseus, 13 fall
between the protostellar and starless completeness limits (similarly
3/20 in Serpens, and 5/17 in Ophiuchus), suggesting that the prestellar 
core lifetime should perhaps be increased by a factor of 1.3.

We must also consider the possible bias introduced by using the number
of embedded protostars, as detected with the higher resolution
\textit{Spitzer} data, rather than the number of Bolocam-detected
protostellar cores, to calculate the prestellar lifetime.
$N_{\mathrm{emb}}/N_{\mathrm{PS}}$ = 1.2, 1.8, and 1.6 in Per, Ser,
and Oph, respectively, as can be seen by comparing
Tables~\ref{coretab} and \ref{lifetab}.  If the starless cores have a
similar "multiplicity" fraction that we missing  due to the
30\arcsec\ resolution of Bolocam, then the prestellar lifetimes would
be increased by these factors.  If cores fragment at later times, or
if multiple protostars collapse from a single core, our lifetimes
would not be affected.

On the other hand, if some fraction of the starless core sample are
\textit{not} prestellar but unbound starless cores, our calculated
lifetime of  $4.5 \times10^5$~yr would be an overestimate of the true
prestellar core lifetime.  Extrapolating from the six cores with
$M_{\mathrm{dust}}/M_{\mathrm{vir}} < 0.5$ in Figure~\ref{virmass},
the lifetime could \textit{decrease} by a factor of 1.2.  Likewise,
the assumed embedded phase lifetime is based on the ratio of Class 0
and Class I protostars to Class II sources; if a large fraction of the
Class I source from Evans et al. (2008, in prep) are not embedded
protostars but edge-on TTuari disks \citep[e.g.][]{crapsi08}, the
embedded phase lifetime could be as low as $2.8 \times 10^5$~yr,
decreasing the prestellar core lifetime by a factor of two.

Finally, the assumption of a simple continuous flow of star
formation for the last 2 Myr may, of course, be incorrect, but without
a detailed star formation rate model (which might be equally
incorrect) it is the best we can do.  We hope to mitigate errors
from variations in the star formation rate by averaging over three
clouds.  Given the range of possible errors, we do not apply
correction factors to the numbers in Table~\ref{lifetab}, but consider
the absolute uncertainty in the prestellar core lifetime to be 
a factor of two in either direction.

\section{Conclusions}\label{discsect}

Utilizing large-scale 1.1~mm surveys \citep{enoch06,young06,enoch07}
together with \textit{Spitzer} IRAC and MIPS maps from the c2d Legacy
program \citep{evans03}, we have carried out an unbiased census of
prestellar and protostellar cores in the Perseus, Serpens, and
Ophiuchus molecular clouds.   We identify a total of 108 starless and
92 protostellar cores in the three cloud sample.  Based on a
comparison of 1.1~mm derived masses to virial masses derived from an
NH$_3$ survey of Perseus cores \citep{ros07}, we conclude that the
majority of our starless cores are likely to be gravitationally bound,
and thus prestellar.

The spatial distributions of both starless and protostellar cores are
similar  in these three molecular with varying  global properties.  In
all three clouds both starless and protostellar cores are found only
at relatively high cloud column densities:  75\% of cores are
associated with $\av \gtrsim 6.5-9.5$~mag in Perseus, $\av \gtrsim
6-10$~mag in Serpens, and $\av \gtrsim 19.5-25.5$~mag in Ophiuchus.
Spatial clustering of starless cores is similar in nature to
protostellar cores but lower in amplitude, based on the two-point
spatial correlation function and peak surface density of cores.  

Cloud environment does appear to have some effect on the physical
properties of starless cores, however, and how they differ from cores
that have already formed protostars.  Starless cores in Perseus are
larger and have lower mean densities than protostellar cores; we
suggest a simple scenario by which protostellar cores might have
evolved from starless cores in that cloud, becoming smaller and denser
at a fixed mass.  In Serpens, it appears that future star formation
will occur in lower mass cores than those that are currently forming
protostars.  Meanwhile, in Ophiuchus we see essentially no difference between
cores that have formed stars and those that have not.  Of the three
clouds, Serpens has the highest mean cloud density (measured within
the $\av=2$ contour; \citealt{enoch07}) and the highest turbulent Mach
number (\citealt{enoch07}; J. Pineda, personal communication), which
may be related to its low fraction of starless cores ($N_{\mathrm{SL}}/N_{\mathrm{emb}}
= 0.4$ compared to 1.0 in the other clouds) and the fact that the
higher mass cores have already formed protostars

The combined prestellar CMD, which includes 108 prestellar cores from
three clouds, has a slope above our completeness limit ($0.8\msun$) of
$\alpha=-2.3\pm0.4$.  This result is consistent with recent
measurements of the stellar initial mass function ($\alpha=-2.3$ to
$-2.8$; e.g. \citealt{reid02}; \citealt{kroupa02}), providing  further
evidence that the final masses of stars are directly linked to the
core formation process.   We place a lower limit on the core collapse
efficiency (the percentage of initial core mass that ends up in the
final star) of 25\%.  A more secure link between the CMD and IMF
requires measurement of the CMD down to masses less than $0.2\msun$
in samples of cores that can be demonstrated
to be likely prestellar.

In all three clouds the lifetime of dense prestellar cores is similar
to the lifetime of embedded protostars,  or $2-5\times10^5$~yr.  The
three-cloud average is $4.5\pm0.8 \times 10^5$~yr  (with an absolute
uncertainty of a factor of two), arguing strongly for dynamic core
evolution on a few free-fall timescales.  Such a short prestellar core
lifetime is inconsistent with highly magnetically sub-critical cores,
in which case evolution should occur over an ambipolar diffusion
timescale ($t_{\mathrm{AD}} \sim 10^7$~yr; \citealt[e.g.,][]{nak98}).
Our results suggest, rather, a dynamic core evolutionary scenario, as
might be appropriate if turbulence dominates the  cloud physics
\citep{mlk04}, or for near-critical magnetic models \citep{cb01}.  The
observed prestellar core lifetime decreases with increasing mean
density,  at a rate faster than that expected from the dependence of
the free-fall  timescale on density.

Although this measurement of the prestellar core lifetime supports a
dynamic paradigm over a quasi-static one, the distinction is not
clear-cut.   Our observations could still be consistent with a
quasi-static picture if we are only observing the densest stages
($n>2\times10^4$~cm$^{-3}$) of a longer-scale core evolution.
Furthermore, the fact that we observe extinction thresholds for
finding dense cores at $\av \gtrsim 6$~mag may be a hint that magnetic
fields become important in the low column density regions of molecular
clouds, inhibiting the formation of high-density prestellar cores (and
thus star formation).    Better measurements of magnetic field
strengths and more detailed comparisons  between observations and
models will be help to resolve these ambiguities.

\acknowledgments

The authors are grateful to Jens Kauffman and Jason Kirk for their
insightful comments and suggestions, and to the anonymous referee for
raising questions and issues that helped to improve this work.
Support for this work, part of the Spitzer Legacy Science Program, was
provided by NASA through contracts 1224608 and 1230782 issued by the
Jet Propulsion Laboratory, California Institute of Technology, under
NASA contract 1407.  Additional support was provided by NASA  through
the Spitzer Space Telescope Fellowship Program and obtained from NASA
Origins Grant NNG04GG24G to the University of Texas at Austin.
Support for the development of Bolocam was provided by NSF grants
AST-9980846 and AST-0206158.  MLE acknowledges support of a Caltech
Moore Fellowship and a Spitzer Space  Telescope Postdoctoral
Fellowship

\clearpage
\LongTables
\vspace{3cm}
\input{tab1.2} 
\clearpage

\input{tab2.2}

\end{document}